\documentclass[11pt]{article}
\usepackage[colorlinks=true,citecolor=blue,linkcolor=blue]{hyperref}
\usepackage{multirow}
\usepackage[left=2cm,right=2cm,top=2cm,bottom=2cm]{geometry}
\usepackage{cite}
\usepackage{psfrag}
\usepackage[demo]{graphicx}
\usepackage{graphicx}
\usepackage{graphics}
\usepackage{epsfig}
\usepackage{booktabs}
\usepackage{amsmath,amsfonts,amssymb}
\usepackage{pstcol,pst-fill,pst-grad}
\usepackage{pstricks,pst-fill,pst-grad}
\usepackage{euscript}
\usepackage{pstricks}
\usepackage{wrapfig}
\usepackage{slashed}
\usepackage[toc,page]{appendix}
\usepackage{here}

\begin{document}
	\title{\vspace{-3cm}
		\hfill\parbox{4cm}{\normalsize \emph{}}\\
		\vspace{1cm}
		{Laser-assisted CP-odd and CP-even Higgs bosons production in THDM}}
	\vspace{2cm}
	
	\author{M. Ouhammou,$^1$ M. Ouali,$^1$ S. Taj,$^1$ R. Benbrik,$^2$ and B. Manaut$^{1,}$\thanks{Corresponding author, E-mail: b.manaut@usms.ma} \\
		{\it {\small$^1$ Polydisciplinary Faculty, Laboratory of Research in Physics and Engineering Sciences,}}\\
		{\it {\small Team of Modern and Applied Physics, Sultan Moulay Slimane University,}}\\
		{\it {\small Beni Mellal, 23000, Morocco.}}\\	
		{\it {\small$^2$ LPFAS, Polydisciplinary Faculty of Safi, UCAM, Morocco}}\\				
	}
	\maketitle \setcounter{page}{1}
\date{}
\begin{abstract}
In this paper, we have theoretically studied the neutral Higgs pair production in Two Higgs Doublet Model (THDM) in the presence of a circularly polarized laser field. The laser-assisted differential partial cross section is derived in the centre of mass frame at the leading order including $Z$ diagram. The total cross section is computed numerically by integrating the differential cross section over the solid angle $d\Omega$. Two benchmark points are discussed for the THDM parameters. In the first step, we have analyzed the total cross section of ${e}^{+}{e}^{-}\rightarrow h^{0}A^{0}$ by considering $H^{0}$ as the standard model-like Higgs boson. Then, the process ${e}^{+}{e}^{-}\rightarrow H^{0}A^{0}$ is studied by taking $h^{0}$ as the Higgs boson of the standard model. For both benchmark points, the laser-assisted total cross section of the studied processes depends on the produced neutral Higgs masses, the centre of mass energy and the laser field parameters. In addition, the maximum cross section occurs at high centre of mass energy for the process ${e}^{+}{e}^{-}\rightarrow H^{0}A^{0}$ as compared to that of ${e}^{+}{e}^{-}\rightarrow h^{0}A^{0}$.
\end{abstract}
\section{Introduction}
The Standard Model of Particle Physics (SM) is tested experimentally, and a beautiful agreement between the theoretical predictions and experimental data has been achieved. The Higgs mechanism is assumed to be the right approach for giving masses to all fermions and gauge bosons through Yukawa coupling terms \cite{glashow}. A signal has already been observed, in 2012, with a mass around 125 GeV \cite{ATLAS,CMS}, and it is believed to be the SM Higgs boson predicted by the Higgs mechanism. However, the SM with one Higgs boson suffers from some limitations. For instance, it does not explain observations like neutrino oscillations and dark matter. Moreover, It does not also provide adequate evidences for matter-antimatter asymmetry of the universe. These constitute strong motivations to look for other theoretical extensions beyond the standard model (BSM). One of those extensions is the popular Two Higgs Doublet Model (THDM) \cite{DELPHI,OPAL}. This BSM has received a great attention mainly because minimal super-symmetry relies on it. In the general THDM, both the scalar Higgs doublets, $\phi_{1}$ and $\phi_{2}$ can couple to fermions of both types, and the Higgs sector contains 5 scalars with at least one neutral Higgs being the SM-like Higgs discovered at the LHC. The remaining Higgs bosons are two charged Higgs bosons, one CP-odd $(A)$ and one CP-even $(H)$.

The first realization of the laser in 1960 \cite{Maiman:1960} is one of the most important technological breakthroughs. Moreover, the rapid development of laser sources in the past decades brings us high-power laser systems. This has been possible mainly due to the continuous progress made along two specific directions: decrease of the laser pulse duration and increase of the laser peak intensity \cite{Mourou and Tajima 2011}. Lasers are indispensable tools for investigating physical processes in different areas ranging from non relativistic atomic and relativistic atomic physic\cite{Manaut1:2004,Taj:2005,Manaut:2004,Manaut1:2005} to nuclear and plasma physics.

Nowadays, such high-power laser systems motivate researchers to study fundamental processes in electroweak and quantum electrodynamics in the presence of a strong electromagnetic field realistic in experiments. Some typical processes in high energy are laser-assisted scattering \cite{Higgs-strahlung,Z-production,muon-pair-prod} and decay processes \cite{Baouahi1,Baouahi2,Jakha}. In \cite{Ouali}, we have discussed the effect of a strong electromagnetic field with a circular polarization on the cross section of the charged Higgs pair production in the Inert Higgs Doublet Model (IHDM). The laser-assisted neutral Higgs pair production in IHDM is discussed in \cite{Neutral-Higgs}, and we have found that the electromagnetic field decreases the cross section. In this respect, we have investigated the same process (${e}^{+}{e}^{-}\rightarrow H^{0}A^{0}$) in THDM in order to analyze the effect of the circularly polarized laser field on its cross section.

The remainder of this research paper is organized as follows: The section 2 is devoted to the theoretical calculation of the differential cross section of the neutral Higgs pair production process ${e}^{+}{e}^{-}\rightarrow A^{0}\,\Phi$ ,\, $( \Phi=h^{0},H^{0}) $, in the presence of a circularly polarized electromagnetic field. The results obtained are discussed in section 3. A short conclusion is given in section 4. In the appendix, we have listed some of the multiplying coefficients of the Bessel function that appear in equation \ref{20}. Throughout this work, we have used natural units such that $(\hbar = c = 1)$. The Livi-Civita tensor is taken as $\varepsilon^{0123}=1$, and the metric $g^{\mu\nu}$ is chosen as $g^{\mu\nu}=(1,-1,-1,-1)$.
\section{Outline of the theory}\label{sec:theory}
\begin{figure}[H]
  \centering
      \includegraphics[width=0.5\textwidth]{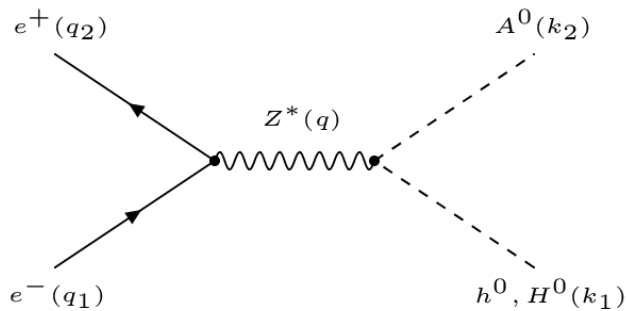}
  \caption{Leading order Feynman diagram for $s$-channel neutral Higgs pair production in the THDM.}
  \label{fig1}
\end{figure}

In this section, we have analytically calculated the differential cross section of the process ${e}^{+}{e}^{-}\rightarrow A^{0}\,\Phi$ ,\, $( \Phi=h^{0},H^{0}) $ in the presence of an electromagnetic field with circular polarization. This processes is described by the Feynman diagram given in figure \ref{fig1}.
The expression of the $Z$-boson couplings to the neutral Higgs bosons are given by the following expressions:
\begin{eqnarray}
A^{0}h^{0}Z^{0}=\nonumber\dfrac{g\cos(\beta -\alpha)}{2\,C_{W}}Z_{\mu}^{0}A^{0}\overleftrightarrow{\partial^{\mu}}h^{0}\quad\text{and}\quad A^{0}H^{0}Z^{0}=-\dfrac{g\sin(\beta -\alpha)}{2\,C_{W}}Z_{\mu}^{0}A^{0}\overleftrightarrow{\partial^{\mu}}H^{0}.
\end{eqnarray}
By using the Feynman rules, the lowest order scattering-matrix element \cite{Greiner:2000} for the laser assisted neutral Higgs-boson pair production can be expressed as follows:
\small
\begin{eqnarray}
S_{fi}({e}^{+}{e}^{-}\rightarrow A^{0}\Phi)&=&-\sin(\beta -\alpha)[\cos(\beta -\alpha)]\frac{-ig}{2C_{W}  }  \int_{}^{} d^4x \int d^4y  \nonumber \\
 &\times &\bar{\psi}_{p_{2},s_{2}}(x)   \Big( \gamma_{\mu} (g_v^{e} -g_a^{e}\gamma^{5}) \Big)   \psi_{p_{1},s_{1}}(x)   D^{\mu\nu}(x-y)  \nonumber \\
 &\times &\phi^{*}_{k_{1}}(y)   \left( \dfrac{ig \overleftrightarrow{\partial_{\nu}}}{2C_{W}} \right)  \phi^{*}_{k_{2}}(y)\qquad\qquad \text{for}\quad \Phi =H^{0}[h^{0}],
 \label{1}
\end{eqnarray}
\normalsize
where $\theta_W$ is the Weinberg angle, $C_{W}=\cos({\theta_{W}})$ and $S_{W}=\sin({\theta_{W}})$.
The electroweak coupling constant {$ g $} is given by $g^{2}=e^{2}/\sin^{2}({\theta_{W}})=8G_{F}M_{Z}^{2}\cos^{2}({\theta_{W}})/\sqrt{2}$, where $e$ is the electron charge.
$g_v^{e}$ and $g_a^{e}$ are the vector and axial vector coupling constants. The matrix $\gamma^{5}$ is expressed in terms of Dirac matrices as follows: $\gamma^{5}=-i\gamma^{0}\gamma^{1}\gamma^{2}\gamma^{3}$. The derivative $\overleftrightarrow{\partial_{\nu}}$ is defined such that: $f\overleftrightarrow{\partial}g=f*\partial g -\partial f*g$, where $f$ and $g$ are a given functions. $x$ and $y$ are the space time coordinates of the incident particles and scattered Higgses, respectively.
Inside the laser field, the electron and positron can be described by Dirac equation as follows:
 \begin{equation}
 \Big[(p_{i}-eA)^{2}-m_{e}^{2}-\dfrac{ie}{2}F_{\mu\nu}\sigma^{\mu\nu} \Big]\psi_{p_{i},s_{i}}(x)=0,
  \label{2}
 \end{equation}
where $p_{i}(i=1, 2)$ and $ s_{i}(i=1, 2)$ refer to the particle momentum and its spin outside the electromagnetic field. $m_{e}$ denotes the mass of the incident particles.
From the equation (2), we can derive the Dirac-Volkov states of the electron and positron as follows \cite{Volkov:1935}:
\begin{equation}
\begin{cases}
\psi_{p_{1},s_{1}}(x)= \Big[1-\dfrac{e \slashed k \slashed A}{2(k.p_{1})}\Big] \frac{u(p_{1},s_{1})}{\sqrt{2Q_{1}V}} \exp^{iS(q_{1},s_{1})}&\\
\psi_{p_{2},s_{2}}(x)= \Big[1+\dfrac{e \slashed k \slashed A}{2(k.p_{2})}\Big] \frac{v(p_{2},s_{2})}{\sqrt{2Q_{2}V}} \exp^{iS(q_{2},s_{2})},
\end{cases}
 \label{3}
\end{equation}
where the argument of the exponential terms, $S(q_{i},s_{i})(i=1,2)$, are expressed as:
\begin{equation}
\begin{cases}
S(q_{1},s_{1})=- q_{1}x +\frac{e(a_{1}.p_{1})}{k.p_{1}}\sin\phi - \frac{e(a_{2}.p_{1})}{k.p_{1}}\cos\phi &\\
S(q_{2},s_{2})=+ q_{2}x +\frac{e(a_{1}.p_{2})}{k.p_{2}}\sin\phi - \frac{e(a_{2}.p_{2})}{k.p_{2}}\cos\phi ,
\end{cases}
 \label{4}
\end{equation}
with, $u(p_{1},s_{1})$ and $v(p_{2},s_{2})$ are the bispinors of the incident charged particles.
$q_{i}=(Q_{i},q_{i})(i=1,2)$ are the effective four-vector momentum of the electron and positron inside the laser field, and $ Q_{i}(i=1,2)$ are their effective energies acquired inside the laser field such that $q_{i}=p_{i}+(e^{2}a^{2}/2(k.p_{i}))k$, with $ q_{i}^{2}=m_{e}^{*^{2}}=(m_{e}^{2}+e^{2}a^{2})$. $\slashed A=\gamma_{\mu}A^{\mu}$, where $A^{\mu}$ is the classical four-vector potential with circular polarization, and it is defined as:
\begin{equation}
A^{\mu}(\phi)=a_{1}^{\mu}\cos\phi+a_{2}^{\mu}\sin\phi,
 \label{5}
\end{equation}
where $ \phi=k.x $ is its phase. The four-vector $ A^{\mu} $ verifies the Lorentz gauge condition $ k^{\mu}A_{\mu}=0 $, with $ k=(\omega,\vec{k}) $ is the wave 4-vector. $ a_{1}^{\mu}=|a|(0,1,0,0) $ and $ a_{2}^{\mu}=\vert a\vert(0,0,1, 0) $ are equal in magnitude such that $ a_{1}^{2}=a_{2}^{2}=a^{2}=-\vert a\vert^{2}=-(\varepsilon_{0}/\omega)^{2} $, and they verify the orthogonality relation $ a_{1}. a_{2}=0 $. $ \varepsilon_{0} $ denotes the strength of the laser field, and $ \omega $ is its frequency.
The vector $\vec{k}$ is chosen to be parallel to the $z$-axis such that $k=(\omega,0,0,\omega)$ and $k^{2}=0$.
The electromagnetic field does not interact with the produced Higgs particles for the fact that they are neutral. Therefore, they may be described by klein-Gordon state given by \cite{Greiner:2000}:
\begin{equation}
\phi_{k_{1}}(y)=\dfrac{1}{\sqrt{2 Q_{A^{0}} V}} e^{-ik_{1}y} \hspace*{0.5cm};\hspace*{0.5cm}  \\\ \phi_{k_{2}}(y)=\dfrac{1}{\sqrt{2 Q_{\Phi} V}} e^{-ik_{2}y},
 \label{6}
\end{equation}
where $k_{1}$ and $k_{2}$ are the four-momentum of the produced neutral Higgs-bosons $A^{0}$ and $\Phi $. $Q_{A^{0}}$ and $Q_{\phi}$ are their quasi-energies.
$ D^{\mu \nu}(x-y) $ is the $Z$-boson propagator \cite{Greiner:2000}, and $q$ is its four vector momentum such that:
 \begin{equation}
D^{\mu\nu}(x-y)=\int \dfrac{d^{4}q}{(2\pi)^4} \frac{e^{-iq(x-y)}}{q^{2}-M_{Z}^{2}}\Bigg[-ig^{\mu\nu}+i\dfrac{q^{\mu}q^{\nu}}{M_{Z}^{2}}\Bigg].
 \label{7}
\end{equation}
By performing the necessary replacements in equation (\ref{1}), the expression of the scattering matrix element becomes as follows:
\begin{eqnarray}
S_{fi}^{n}({e}^{+}{e}^{-}\rightarrow A^{0}\Phi)&=&-\sin(\beta -\alpha)[\cos(\beta -\alpha)]\dfrac{1}{4V^{2}\sqrt{Q_{1}Q_{2}Q_{A^{0}}Q_{\Phi}}}\nonumber \\
 &\times &\left(\dfrac{g^{2}}{4C_{W}^{2}}\right)  \Bigg( \dfrac{1}{(q_{1}+q_{2}+nk)^{2}-M_{Z}^{2}}  \Bigg) \nonumber \\
 &\times &  \bar{v}(p_{2},s_{2})\Gamma^{n}u(p_{1},s_{1}) \qquad\qquad \text{for}\quad \Phi =H^{0}[h^{0}].
  \label{8}
\end{eqnarray}
The number of transferred photons $n$ between the electromagnetic field and the colliding electron-positron may be positive in the case of absorption or negative in the case of emission. The quantity $\Gamma^{n}_{\mu}$ is given by the following equation:
\begin{equation}
 \Gamma^{n}=\chi^{0}\,A_{0n}(z)+\chi^{1}\,A_{1n}(z)+\chi^{2}\,A_{2n}(z),
  \label{9}
 \end{equation}
where $\chi^{0}$, $\chi^{1}$ and $\chi^{2}$ can be expressed as follows:
\begin{equation}
\begin{cases}\chi^{0}=(k_{1}^{\mu}-k_{2}^{\mu})\gamma_{\mu}(g_{v}^{e}-g_{a}^{e}\gamma^{5})+2\rho_{p_{1}}\rho_{p_{2}}a^{2}(k_{1}^{\mu}-k_{2}^{\mu})k_{\mu}\slashed k(g_{v}^{e}-g_{a}^{e}\gamma^{5})   &\\
\chi^{1}=\rho_{p_{1}}(k_{1}^{\mu}-k_{2}^{\mu})\gamma_{\mu}(g_{v}^{e}-g_{a}^{e}\gamma^{5})\slashed k\slashed a_{1}-\rho_{p_{2}}(k_{1}^{\mu}-k_{2}^{\mu})\slashed a_{1}\slashed k \gamma_{\mu}(g_{v}^{e}-g_{a}^{e}\gamma^{5})   &\\
\chi^{2}=\rho_{p_{1}}(k_{1}^{\mu}-k_{2}^{\mu})\gamma_{\mu}(g_{v}^{e}-g_{a}^{e}\gamma^{5})\slashed k\slashed a_{2}-\rho_{p_{2}}(k_{1}^{\mu}-k_{2}^{\mu})\slashed a_{2}\slashed k \gamma_{\mu}(g_{v}^{e}-g_{a}^{e}\gamma^{5}) \end{cases},
\label{10}
\end{equation}
with $\rho_{p_{i}}=e/2(kp_{i})$,$ (i=1,2) $.
The quantities $A_{0n}(z)$, $A_{1n}(z)$ and $A_{2n}(z)$ can be expressed in terms of Bessel functions as follows:
\begin{equation}
\left.
  \begin{cases}
     A_{0n}(z) \\
      A_{1n}(z) \\
      A_{2n}(z)
  \end{cases}
  \right\} = \left.
  \begin{cases}
     J_{n}(z)e^{-in\phi _{0}}\\
    \frac{1}{2}\Big(J_{n+1}(z)e^{-i(n+1)\phi _{0}}+J_{n-1}(z)e^{-i(n-1)\phi _{0}}\Big) \\
     \frac{1}{2\, i}\Big(J_{n+1}(z)e^{-i(n+1)\phi _{0}}-J_{n-1}(z)e^{-i(n-1)\phi _{0}}\Big)
  \end{cases}
  \right\}.
  \label{11}
\end{equation}
In equation (\ref{11}), the argument of the Bessel function $z$ and its phase $\phi_{0}$ are given by:
$ z=\sqrt{\xi_{1}^{2}+\xi_{2}^{2}}$ and $\phi_{0}= \arctan(\xi_{2}/\xi_{1})$, where:
\begin{center}
$\xi_{1}=\dfrac{e(a_{1}.p_{1})}{(k.p_{1})}-\dfrac{e(a_{1}.p_{2})}{(k.p_{2})}$ \qquad ; \qquad $\xi_{2}=\dfrac{e(a_{2}.p_{1})}{(k.p_{1})}-\dfrac{e(a_{2}.p_{2})}{(k.p_{2})}$.
\end{center}
In the centre-of-mass frame, the differential cross section can be expressed as follows\cite{Greiner:2000}:
\begin{equation}
d\sigma^{n}=\dfrac{|S_{fi}^{n}|^{2}}{VT}\frac{1}{|J_{inc}|}\frac{1}{\varrho}V\int_{}\dfrac{d^{3}k_{1}}{(2\pi)^3}V\int_{}\dfrac{d^{3}k_{2}}{(2\pi)^3},
\label{12}
\end{equation}
where $\varrho=V^{-1}$ denotes the particle density, and $|J_{inc}|=(\sqrt{(q_{1}q_{2})^{2}-m_{e}^{*^{4}}}/{Q_{1}Q_{2}V})$ is the current of incident particles in the centre of mass frame.
The spin-unpolarized differential cross section is obtained by averaging over the initial spins and summing over the final ones. Thus, we obtain after calculations that:
\small
\begin{eqnarray}
d\bar{\sigma}({e}^{+}{e}^{-}\rightarrow A^{0}\Phi)&=&\sin^{2}(\beta -\alpha)[\cos^{2}(\beta -\alpha)]\dfrac{g^{4}}{128C_{W}^{4}}  \Bigg[ \dfrac{1}{(q_{1}+q_{2}+nk)^{2}-M_{Z}^{2}}  \Bigg]^{2} \nonumber \\
 &\times & \dfrac{1}{\sqrt{(q_{1}q_{2})^{2}-m_{e}^{*^{4}}}} \big|\overline{M_{fi}^{n}} \big|^{2}   \int \dfrac{|\mathbf{k}_{1}|^{2}d|\mathbf{k}_{1}|d\Omega}{(2\pi)^{2}Q_{\phi}}   \\
 &\times &\nonumber   \int \dfrac{d^{3}k_{2}}{Q_{A^{0}}} \delta^{4}(k_{1}+k_{2}-q_{1}-q_{2}-nk)  \qquad\qquad \text{for}\quad \Phi =H^{0}[h^{0}].
\label{13}
\end{eqnarray}
\normalsize
The remaining integral over $d^{3}k_{1}$ is calculated by using the well-known formula \cite{Greiner:2000} given by:
\begin{equation}
 \int d\mathbf y f(\mathbf y) \delta(g(\mathbf y))=\dfrac{f(\mathbf y)}{|g^{'}(\mathbf y)|_{g(\mathbf y)=0}}.
 \label{14}
\end{equation}
Thus, the expression of the differential cross section becomes as follows:
\begin{eqnarray}
\dfrac{d\bar{\sigma}}{d\Omega}({e}^{+}{e}^{-}\rightarrow A^{0}\Phi)&=&\sin^{2}(\beta -\alpha)[\cos^{2}(\beta -\alpha)]\dfrac{g^{4}}{128C_{W}^{4}} \Bigg[ \dfrac{1}{(q_{1}+q_{2}+nk)^{2}-M_{Z}^{2}}  \Bigg]^{2} \nonumber \\
 &\times &  \dfrac{1}{\sqrt{(q_{1}q_{2})^{2}-m_{e}^{*^{4}}}}  \big|\overline{M_{fi}^{n}} \big|^{2}\dfrac{2|\mathbf{k_{1}}|^{2}}{(2\pi)^{2}Q_{\phi}} \nonumber \\
 &\times &  \dfrac{1}{\big|g^{'}(|\mathbf{k}_{1}|)\big|_{g(|\mathbf{k}_{1}|)=0}}\qquad\qquad \text{for}\qquad \Phi =H^{0}[h^{0}],
\label{15}
\end{eqnarray}
where the function $g^{'}(|\mathbf{k}_{1}|)$ is given by:
\begin{equation}
 g^{'}(|\mathbf{k}_{1}|)=\dfrac{4 e^{2}a^{2}}{\sqrt{s}}\dfrac{|\mathbf{k}_{1}|}{\sqrt{|\mathbf{k}_{1}|^{2}+M_{\phi}^{2}}}-\dfrac{2|\mathbf{k}_{1}|(\sqrt{s}+n\omega)}{\sqrt{|\mathbf{k}_{1}|^{2}+M_{\phi}^{2}}}.
 \label{16}
\end{equation}
In equation (\ref{15}), the term $ \big|\overline{M_{fi}^{n}} \big|^{2}$ can be expressed as follows:
\begin{equation}
\big|\overline{M_{fi}^{n}} \big|^{2}=\dfrac{1}{4}\sum_{n=-\infty}^{+\infty}\sum_{s}\big|M_{fi}^{n} \big|^{2}=\frac{1}{4}\sum_{n=-\infty}^{+\infty}Tr\Big[ (\slashed p_{2}+m_{e})\Gamma^{n}(\slashed p_{1}-m_{e}) \bar{\Gamma}^{n}\Big],
\label{17}
\end{equation}
where the quantity $\Gamma^{n}$ is given by equation (\ref{9}), and $\bar{\Gamma}^{n}$ is expressed by:
\begin{equation}
\bar{\Gamma}^{n}=\bar{\chi}^{0}\,A_{0n}^{*}(z)+\bar{\chi}^{1}\,A_{1n}^{*}(z)+\bar{\chi}^{2}\,A_{2n}^{*}(z),
 \label{18}
 \end{equation}
with
\small
\begin{equation}
\begin{cases}\bar{\chi}^{0}=(k_{1}^{\nu}-k_{2}^{\nu})\gamma_{\nu}(g_{v}^{e}-g_{a}^{e}\gamma^{5})+2\rho_{p_{1}}\rho_{p_{2}}a^{2}(k_{1}^{\nu}-k_{2}^{\nu})k_{\nu}\slashed k(g_{v}^{e}-g_{a}^{e}\gamma^{5})   &\\
\bar{\chi}^{1}=\rho_{p_{1}} \slashed a_{1} \slashed k (k_{1}^{\nu}-k_{2}^{\nu})\gamma_{\nu}(g_{v}^{e}-g_{a}^{e}\gamma^{5}) -\rho_{p_{2}}(k_{1}^{\nu}-k_{2}^{\nu})\gamma_{\nu}(g_{v}^{e}-g_{a}^{e}\gamma^{5})\slashed k\slashed a_{1}   &\\
\bar{\chi}^{2}=\rho_{p_{1}} \slashed a_{2} \slashed k (k_{1}^{\nu}-k_{2}^{\nu})\gamma_{\nu}(g_{v}^{e}-g_{a}^{e}\gamma^{5}) -\rho_{p_{2}}(k_{1}^{\nu}-k_{2}^{\nu})\gamma_{\nu}(g_{v}^{e}-g_{a}^{e}\gamma^{5})\slashed k\slashed a_{2}. \end{cases}
\label{19}
\end{equation}
\normalsize
In equation (\ref{17}), the trace calculation leads to the following result:
 \small
\begin{eqnarray}
\big|\overline{M_{fi}^{n}} \big|^{2}&=&\frac{1}{4}\sum_{n=-\infty}^{+\infty}\Big[ \Delta_{1}J_{n}^{2}(z)+\Delta_{2}J_{n+1}^{2}(z)+\Delta_{3}J_{n-1}^{2}(z)+\Delta_{4}J_{n}(z)J_{n+1}(z)+\Delta_{5}J_{n}(z)J_{n-1}(z)\nonumber \\
 &+ & \Delta_{6}J_{n-1}(z)J_{n+1}(z) \Big],
\label{20}
\end{eqnarray}
\normalsize
where the quantities $\Delta_{i}$ $(i=1,2,3,4,5\, \text{and}\, 6)$ are calculated using FeynCalc program \cite{FeynCalc}. We give here the expression of the coefficient multiplied by $J_{n}^{2}(z)$, while the other terms are listed in the appendix.
\small
\begin{eqnarray}
\Delta_{1}&=&\nonumber\dfrac{1}{(k.p_{1})(k.p_{2})}\Big[ 2 (a^4 e^4 (g_{a}^{e^{2}} + g_{v}^{e^{2}}) ((k.k_{1}) - (k.k_{2}))^2 +
    2 a^2 e^2 ((k.k_{1}) - (k.k_{2})) (g_{a}^{e^{2}} (((k.k_{1}) - (k.k_{2}))\\&\times &\nonumber (m_{e}^{2} - (p_{1}.p_{2})) + (k.p_{2}) (p_{1}.k_{1}) - (k.p_{2}) (p_{1}.k_{2}) +
          (k.p_{1}) (p_{2}.k_{1}) - (k.p_{1}) (p_{2}.k_{2})) + g_{v}^{e^{2}} (-((k.k_{1}) \\&- &\nonumber (k.k_{2})) (m_{e}^{2} + (p_{1}.p_{2})) + (k.p_{2}) (p_{1}.k_{1}) - (k.p_{2}) (p_{1}.k_{2}) + (k.p_{1}) (p_{2}.k_{1}) - (k.p_{1}) (p_{2}.k_{2}))) \\&+ &\nonumber
    2 (k.p_{1}) (k.p_{2}) (2 g_{a}^{e^{2}} (((p_{1}.k_{1}) - (p_{1}.k_{2})) ((p_{2}.k_{1}) - (p_{2}.k_{2})) + (m_{e}^{2} -
             (p_{1}.p_{2})) (M_{A^{0}}^{2} - (k_{1}.k_{2}))) \\&- &\nonumber
       g_{v}^{e^{2}} (m_{e}^{2} M_{\Phi}^{2} + M_{A^{0}}^{2} (m_{e}^{2} + 2 (p_{1}.p_{2})) -
          2 ((p_{1}.k_{1}) - (p_{1}.k_{2})) ((p_{2}.k_{1}) - (p_{2}.k_{2})) \\&- &2 (m_{e}^{2} + (p_{1}.p_{2})) (k_{1}.k_{2}))))\Big]\qquad\qquad\qquad \text{for}\qquad \Phi =H^{0}[h^{0}].
\end{eqnarray}
We recall that the gauge invariance is well verified for the two cases: Landau gauge $\xi = 0$ and Feynman gauge $\xi = 1$. In all calculations, the Feynman gauge is used.
\normalsize
\section{Results and Discussion}
In this section, we will analyze and discuss the results obtained about the behavior of the total cross section of neutral Higgs pair production process in THDM inside an electromagnetic field with circular polarization. The total cross section is obtained by performing a numerical integration of the differential cross section given by equation (\ref{15}) over the solid angle $d\Omega=\sin\theta d\theta d\phi$. In addition, the total cross section is summed over the number of exchanged photons from $-n$ to $+n$ to obtain the laser-assisted total cross section. Based on its analytical expression, we remark that the cross section depends on the collider centre of mass energy, the produced Higgs masses and the laser parameters such as number of transferred photons, laser field strength and its frequency.
We have taken the parameters of the standard model from PDG \cite{PDG:2021} such that: $m_{e} = 0.511\, MeV$, $m_{Z}=91.1875\,GeV$, the mixing angle $\sin^{2}(\theta_{W})=0.23126$ and the Fermi coupling constant $G_{F}=1.166 3787 \times 10^{-5}\, GeV^{-2}$.
\subsection{The process ${e}^{+}{e}^{-}\rightarrow A^{0}h^{0}$}
In this part, we consider the process $e^{+}e^{-}\rightarrow A^{0}h^{0}$ where the other  scalar Higgs, $H^{0}$, is chosen as being the standard model-like Higgs boson with a mass $125\,GeV$.
The benchmark points chosen for the  produced Higgs boson masses are $M_{A^{0}}=169\,GeV$ and $M_{h^{0}}=95\,GeV$, and $\sin(\beta-\alpha)$ is taken as $\sin(\beta-\alpha)=-0.006$ \cite{Benbrik}.
Let's begin our discussion by analyzing the behavior of the partial total cross section inside the electromagnetic field.
\begin{figure}[H]
  \centering
      \includegraphics[scale=0.58]{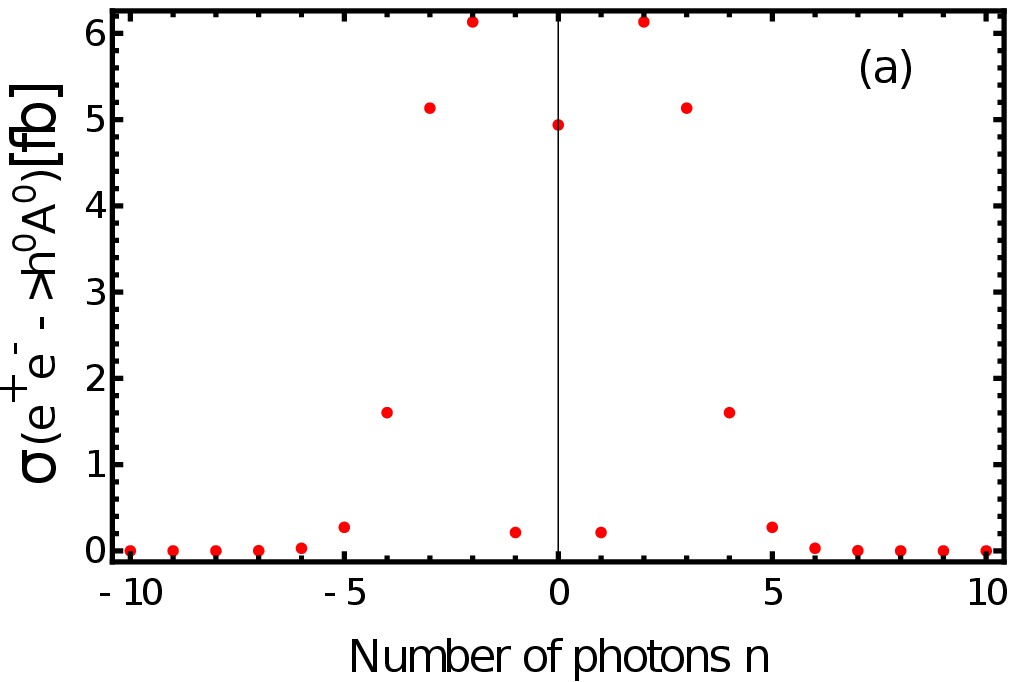}\hspace*{0.4cm}
      \includegraphics[scale=0.59]{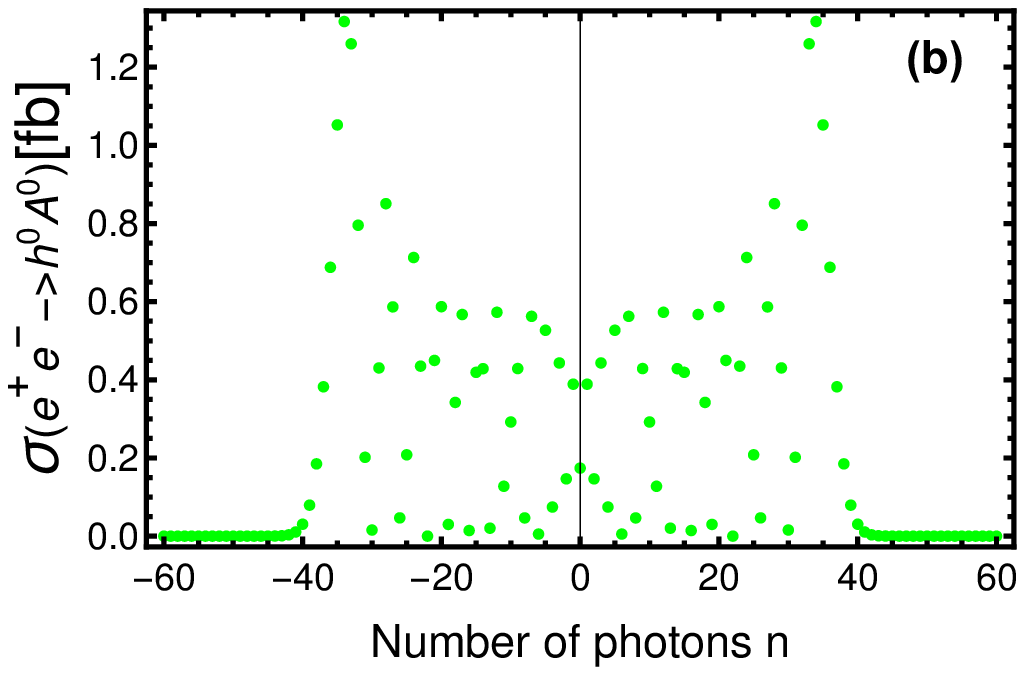}\par\vspace*{0.5cm}
      \includegraphics[scale=0.58]{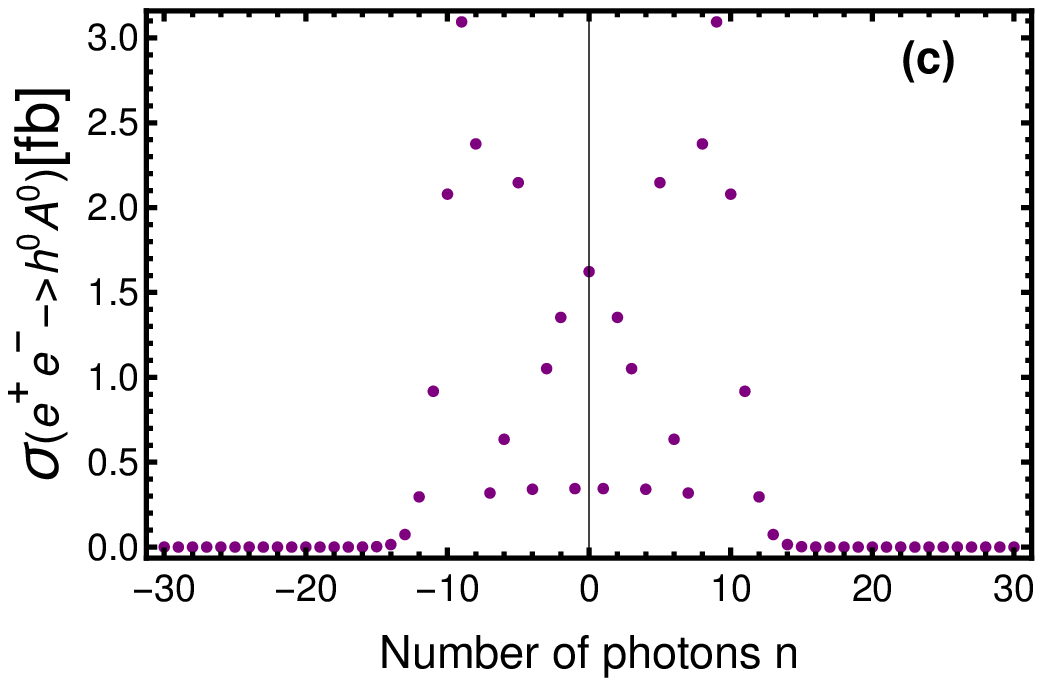}\hspace*{0.4cm}
      \includegraphics[scale=0.58]{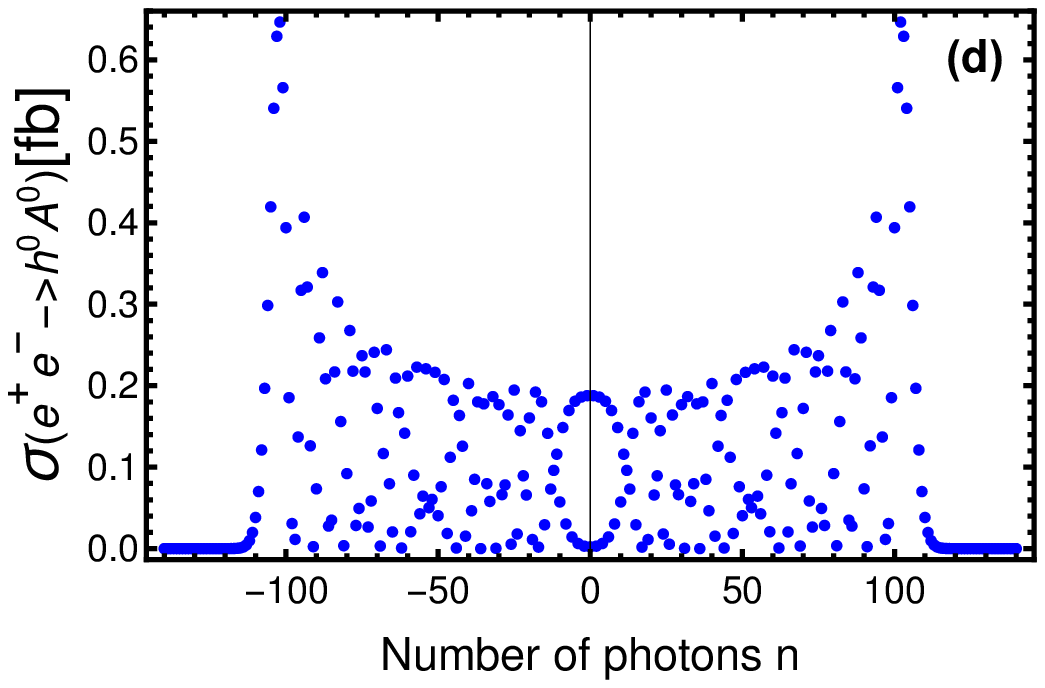}
        \caption{Dependence of the laser-assisted partial total cross section of $ {e}^{+}{e}^{-}\rightarrow A^{0}h^{0} $ on the number of exchanged photons by taking $M_{A^{0}}=169\,GeV$ and $M_{h^{0}}=95\,GeV$ . The laser field strength and its frequency are taken as: $\varepsilon_{0}=10^{5}\,V.cm^{-1}$ and $\omega=2\, eV$ (He:Ne Laser) in (a);  $\varepsilon_{0}=10^{6}\,V.cm^{-1}$ and $\omega=2\, eV$ in (b);  $\varepsilon_{0}=10^{5}\,V.cm^{-1}$ and $\omega=1.17\, eV$ (Nd:YAG laser) in (c);  $\varepsilon_{0}=10^{6}\,V.cm^{-1}$ and $\omega=1.17\, eV$ in (d).}
        \label{fig2}
\end{figure}
Figure \ref{fig2} represents the variation of the partial total cross section as a function of the number of exchanged photons for different laser field strengths and frequencies. The laser field strength is chosen as $\varepsilon_{0}=10^{5}\,V.cm^{-1}$ in  figure \ref{fig2}(a) and \ref{fig2}(c), and $\varepsilon_{0}=10^{6}\,V.cm^{-1}$ in figures \ref{fig2}(b) and \ref{fig2}(d). Figures \ref{fig2}(a) and \ref{fig2}(b) are obtained by using the \textbf{He:Ne laser} which delivers radiations with the frequency $\omega=2\, eV$. The \textbf{Nd:YAG laser} with the frequency $\omega=1.17\, eV$ is used in figures \ref{fig2}(c) and \ref{fig2}(d).
Regardless of the laser parameters values, we observe that the partial total cross section, in each case, presents two symmetric cutoffs. In addition, the partial total cross section which corresponds to the emission of photons ($+n$) is equal to that corresponding to the absorption ($+n$). From one hand, by comparing figures \ref{fig2}(a) and \ref{fig2}(b), we notice that, for the same value of the laser frequency ($\omega=2\, eV$), the maximal number of photons that can be exchanged (cutoff) increases by increasing the laser field strength. For instance, this cutoff is equal to $\pm6$ for $\varepsilon_{0}=10^{5}\,V.cm^{-1}$, and it is equal to $\pm40$ for $\varepsilon_{0}=10^{6}\,V.cm^{-1}$. From the other hand, if we compare results of figure \ref{fig2}(a) to that of figure \ref{fig2}(c), for the same laser field strength ($\varepsilon_{0}=10^{5}\,V.cm^{-1}$), we remark that the cutoff increases so far as the frequency is decreased. For example, cutoffs in these figures are $\pm6$ and $\pm15$ for $\omega=2\, eV$ and $\omega=1.17\, eV$, respectively. Consequently, the maximal number of photons that can be exchanged between the electromagnetic field and the colliding physical system increases either by increasing the laser field strength or by decreasing its frequency. Let's move, now, to discuss the behavior of the summed total cross section inside an electromagnetic field with a circular polarization.
\begin{figure}[H]
  \centering
      \includegraphics[scale=0.58]{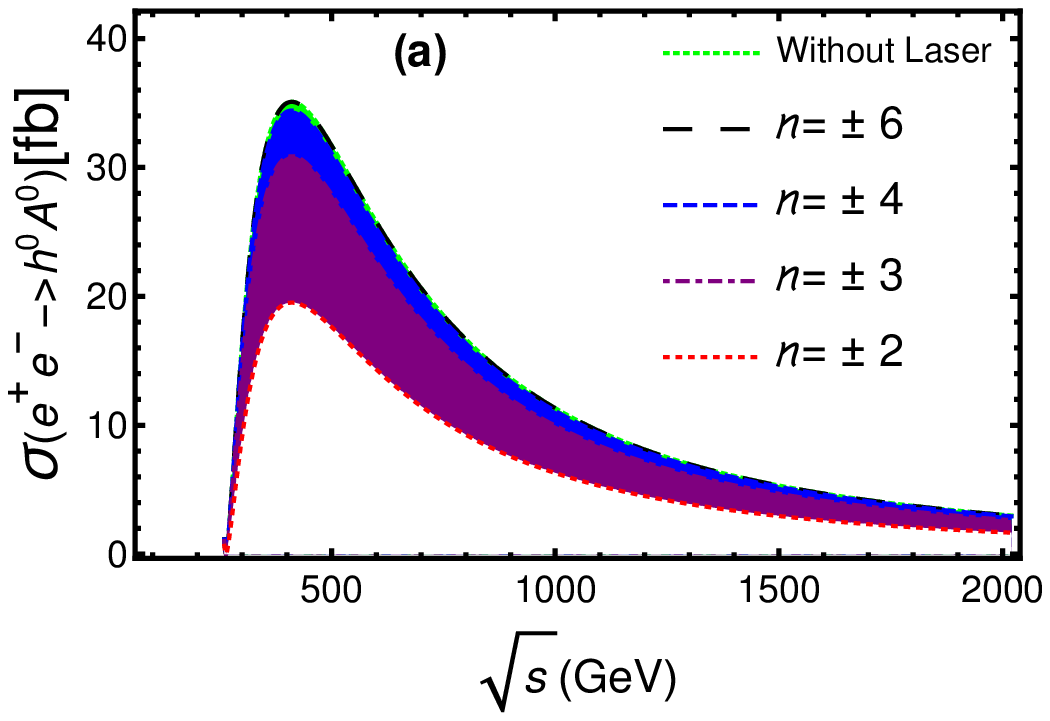}\hspace*{0.4cm}
      \includegraphics[scale=0.58]{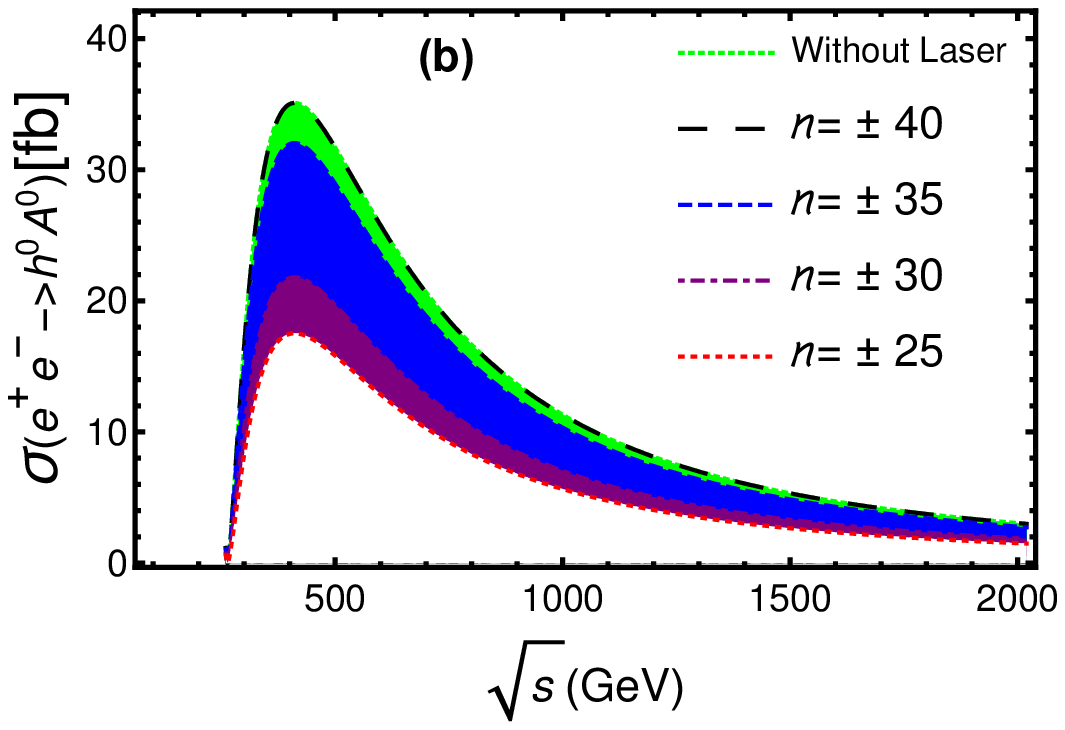}\par\vspace*{0.5cm}
      \includegraphics[scale=0.57]{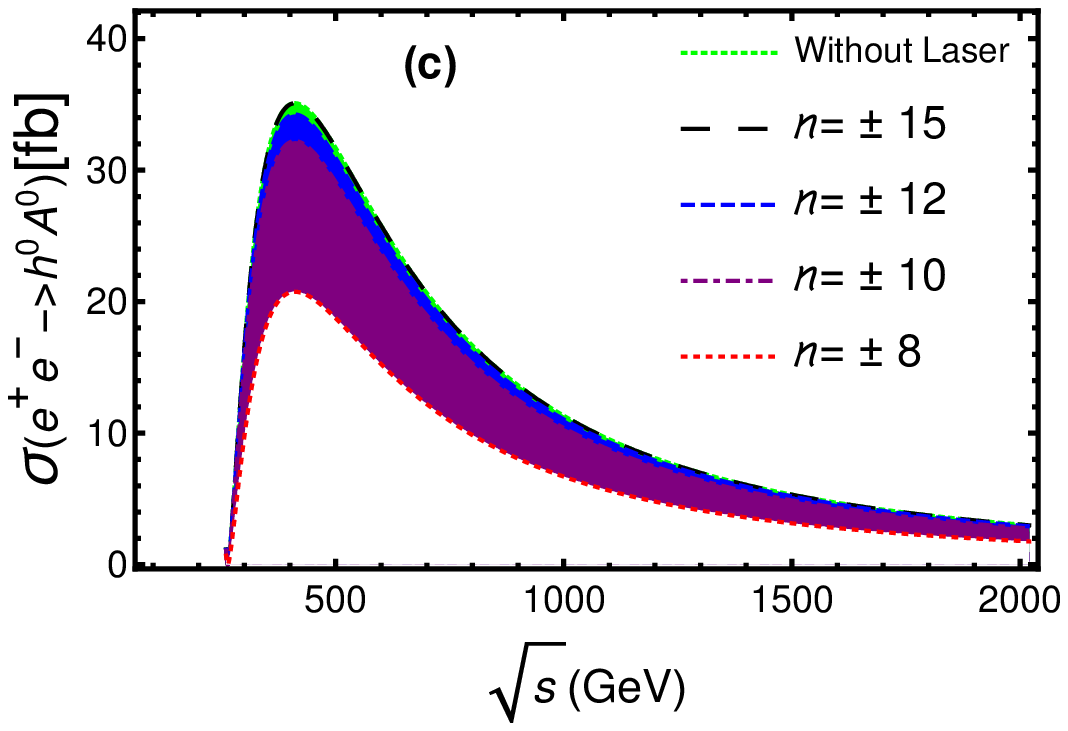}\hspace*{0.4cm}
      \includegraphics[scale=0.58]{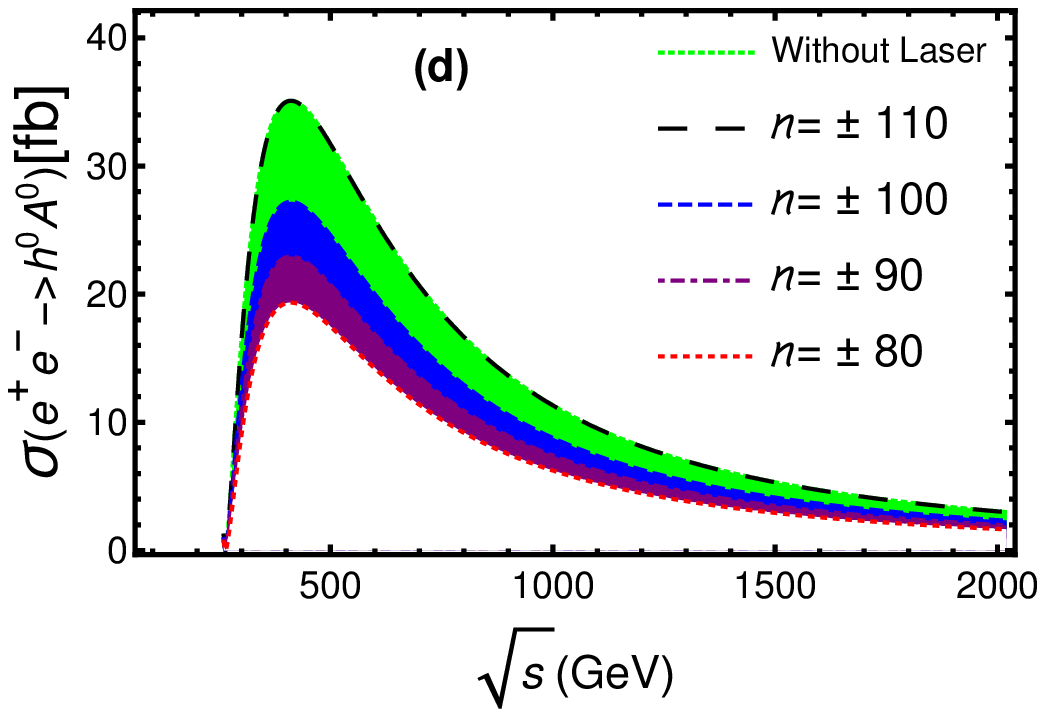}
        \caption{Variation of the laser-assisted total cross section of the process ${e}^{+}{e}^{-}\rightarrow A^{0}h^{0}$ as a function of the centre of mass energy for different number of exchanged photons and by taking $M_{A^{0}}=169\,GeV$ and $M_{h^{0}}=95\,GeV$. The frequency is taken as $\omega=2\, eV$ (\textbf{He:Ne Laser}) in (a) and (b), and $\omega=1.17\, eV$ (\textbf{Nd:YAG laser}) in (c) and (d). The laser field strength is chosen as $\varepsilon_{0}=10^{5}\,V.cm^{-1}$ in (a) and (c), and $\varepsilon_{0}=10^{6}\,V.cm^{-1}$ in (b) and (d).}
        \label{fig3}
\end{figure}
Figure \ref{fig3} illustrates the variation of the total cross section as a function of the centre of mass energy for different number of exchanged photons, laser field strengths and frequencies. Let's start by discussing the behavior of the laser-free cross section. The latter increases in a small range of low centre of mass energies, and its maximum is approximately reached at $\sqrt{s}=410\, GeV$. Beyond this maximum, the cross section decreases progressively as far as the centre of mass energy increases, and this is due to the phase space suppression. Inside the electromagnetic field, the total cross section has a similar aspect as its corresponding laser-free total cross section. However, its order of magnitude depends on the number of transferred photons between the laser beam and the ${e}^{+}{e}^{-}$ beam. More precisely, regardless of the laser field strength and its frequency, the total cross section raises by increasing the number of exchanged photons. In addition, for a certain number of photons exchanged, the laser-assisted total cross section will have the same order as its corresponding laser-free cross section. The required number of photons to be transferred in order to reach the laser free cross section strongly depends on the laser field strength and its frequency. For instance, this number is $\pm6$, $\pm40$, $\pm15$ and $\pm110$ in figures \ref{fig3}(a), \ref{fig3}(b), \ref{fig3}(c) and \ref{fig3}(d), respectively. We remark that this numbers correspond to the cutoffs obtained in figure \ref{fig2}.
By comparing figures \ref{fig3}(a) and \ref{fig3}(b), for the same laser frequency ($\omega=1.17\, eV$), the maximum number of photons that can be exchanged increases by increasing the laser field strength. In addition, for the same laser field strength as in figures \ref{fig3}(a) and \ref{fig3}(c), the cutoff raises by decreasing the laser field frequency. For a given laser field strength and frequency, the summation over its corresponding cutoffs leads to the well known sum-rule which is elaborated by Kroll and Watson in \cite{Kroll+Watson}. Since we have considered in this section that $H^{0}$ is the standard model-like Higgs boson, the neutral Higgs bosons $h^{0}$ and $A^{0}$ have no precise mass. Therefore, it is important to study their mass spectrum inside the laser field.
Figure \ref{fig4} represents the variation of the total cross section as a function of the CP-even Higgs boson mass for different laser parameters such as the number of exchanged photons and laser field frequency. The CP-odd Higgs mass and the laser field strength are successively chosen as $M_{A^{0}}=169\,GeV$ and $\varepsilon_{0}=10^{5}\,V.cm^{-1}$.
\begin{figure}[H]
  \centering
      \includegraphics[scale=0.58]{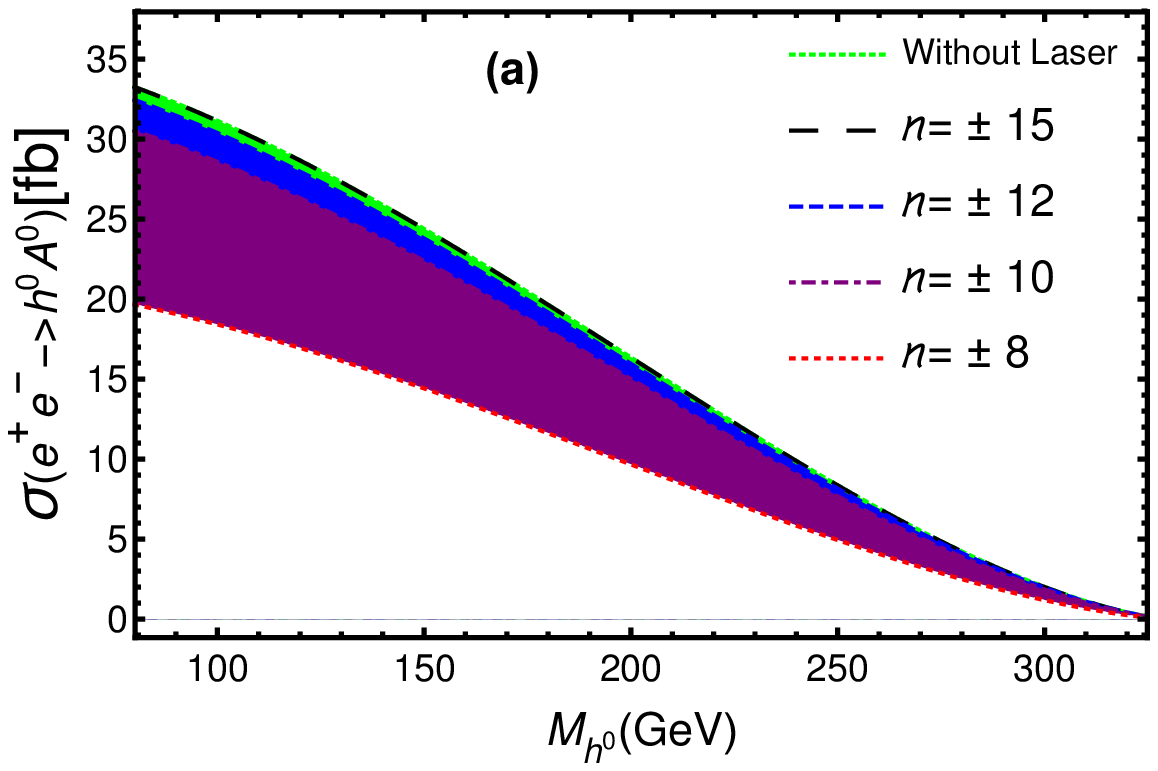}\hspace*{0.4cm}
      \includegraphics[scale=0.58]{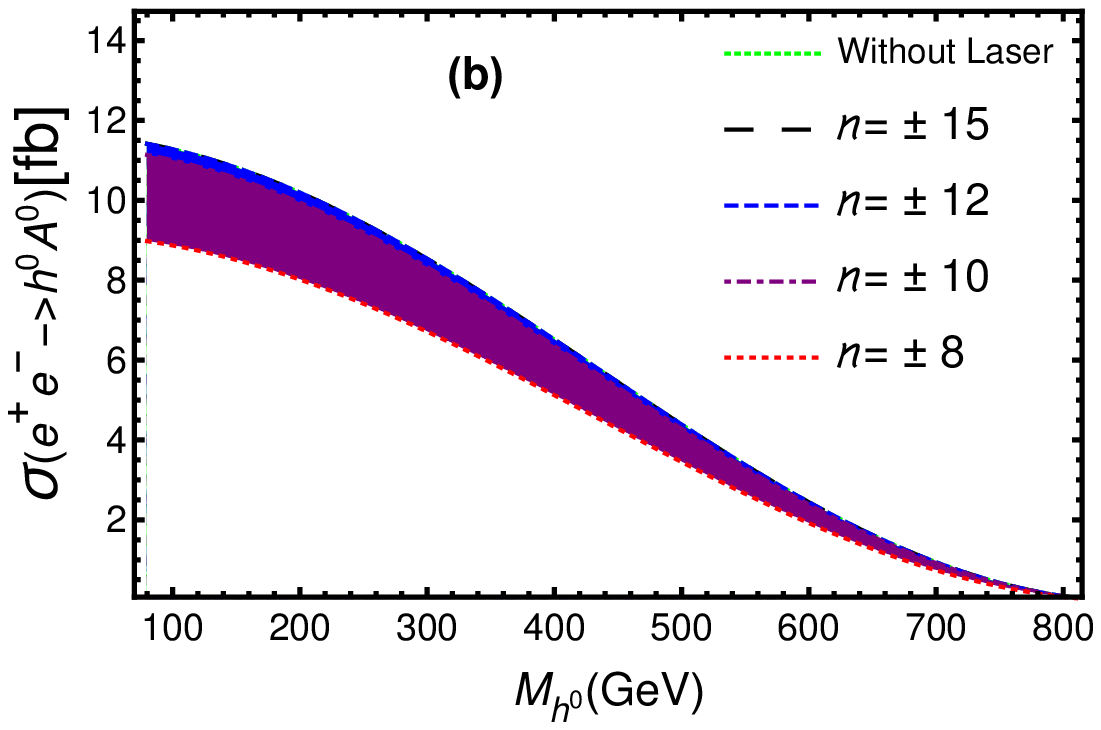}\par\vspace*{0.5cm}
      \includegraphics[scale=0.58]{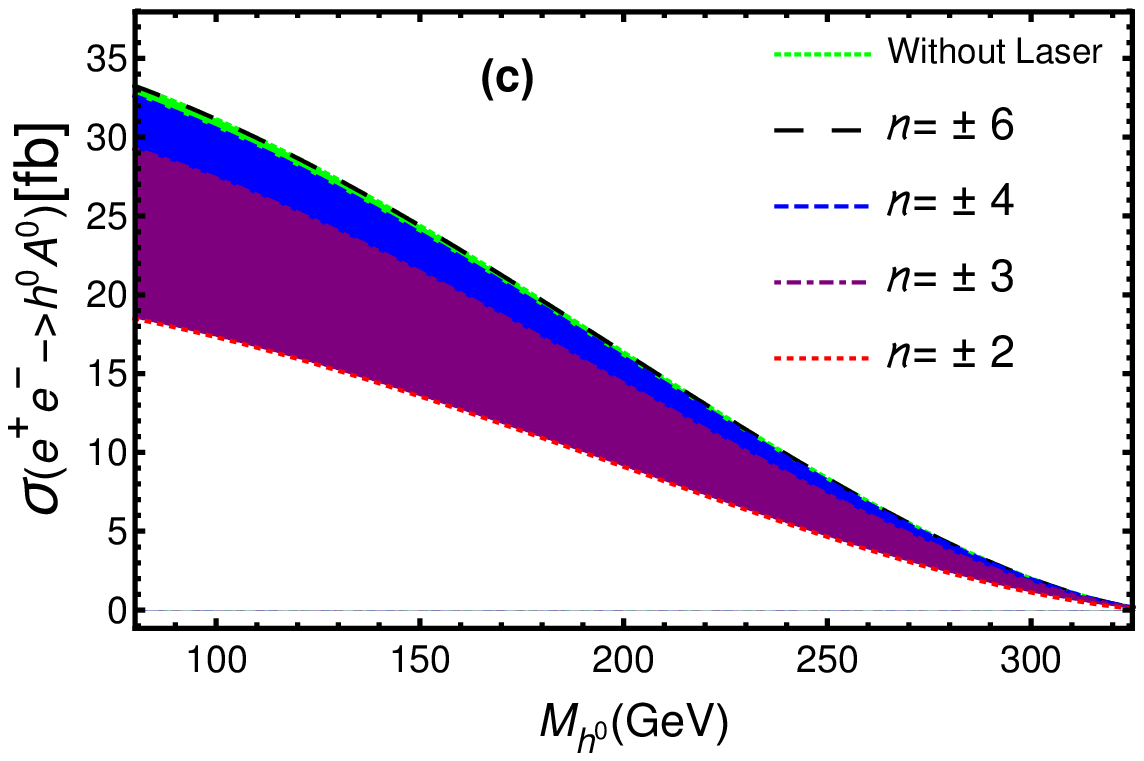}\hspace*{0.4cm}
      \includegraphics[scale=0.6]{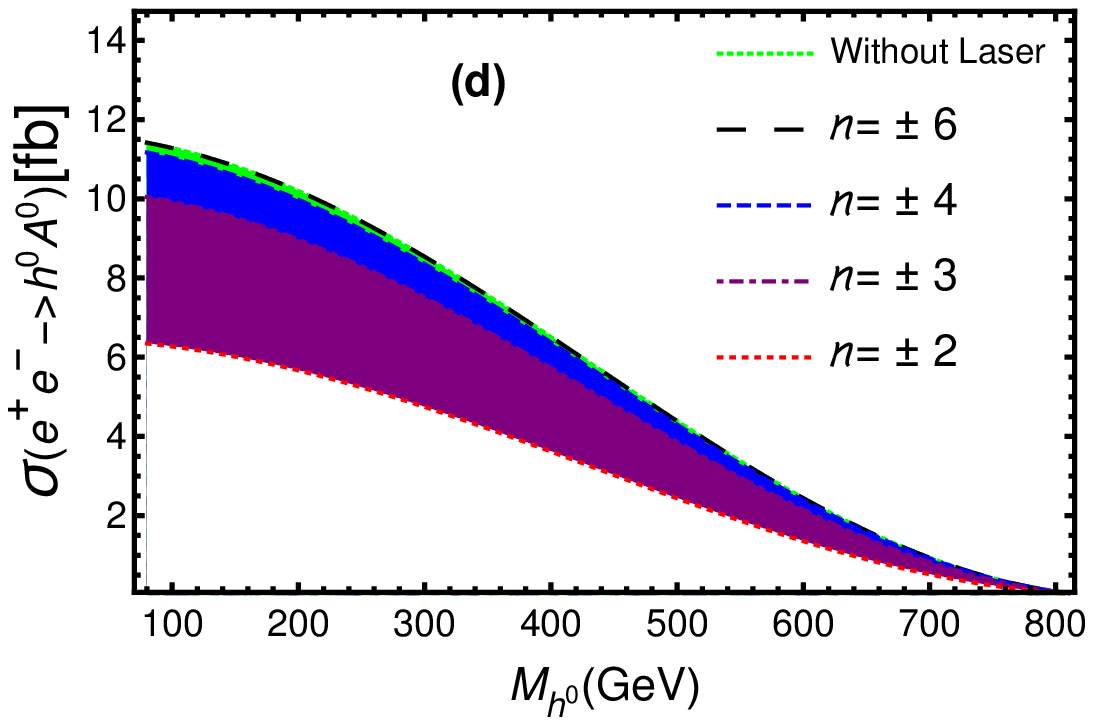}
        \caption{Dependence of the laser-assisted total cross section of the process ${e}^{+}{e}^{-}\rightarrow h^{0}A^{0}$ on the CP-even Higgs-boson mass for different number of exchanged photons and for two centre of mass energies:  $500\,GeV$(left panel) and $1\,TeV$ (right panel). The CP-odd Higgs mass and the laser field strength are chosen as $M_{A^{0}}=169\,GeV$ and  $\varepsilon_{0}=10^{5}\,V.cm^{-1}$, respectively. The \textbf{Nd:YAG laser} $(\omega=1.17\, eV)$ is used in (a) and (b), while in (c) and (d), the \textbf{He:Ne laser} $(\omega=1.17\, eV)$ is used.}
\label{fig4}
\end{figure}
The laser-free total cross section has its maximum at low CP-even Higgs masses, and it decreases as far as the mass increases until it becomes null at $M_{h^{0}}=320\,GeV$ and $M_{h^{0}}=800\,GeV$ for the centre of mass energies $\sqrt{s}= 500\,GeV$ and $\sqrt{s}=1\,TeV$, respectively. However, its order of magnitude strongly depends on the collider centre of mass energy. For instance, for the right panel in which $\sqrt{s}=500\,GeV$, the maximum cross section is approximately $\sigma_{max}=32\,[fb]$. Whereas,  $\sigma_{max}=11.33\,[fb]$ for $\sqrt{s}=1\,TeV$ (left panel). For the laser-assisted total cross section, it is obvious from these figures that it has the same general aspect as its corresponding laser-free cross section, yet its order of magnitude depends also on the CP-even Higgs mass and the centre of mass energy in addition to the number of exchanged photons. To be more precise, it increases as much as we increase the number of exchanged photons until it reaches its corresponding laser-free total cross section. For example, its values in figure \ref{fig4}(a) are successively equal to $\sigma=18.68\,[fb]$, $\sigma=28.5\,$[fb] and $\sigma=30.43\,[fb]$ for $n=\pm8$, $n=\pm10$ and $n=\pm12$.
Another important remark is that for the same number of transferred photons, the cross section differs from one frequency to another and also from one strength to another. For instance, for $n=\pm8$ the cross section is $\sigma=18.68\,[fb]$ and $\sigma=8.96\,[fb]$ in figure \ref{fig4}(a) and \ref{fig4}(b), respectively. In addition, the laser-assisted total cross section decreases by increasing the CP-even Higgs mass. We have made, in figure \ref{fig5}, the same analysis for the case where the CP-even Higgs mass is fixed as $M_{h^{0}}=95\,GeV$.
\begin{figure}[H]
  \centering
      \includegraphics[scale=0.6]{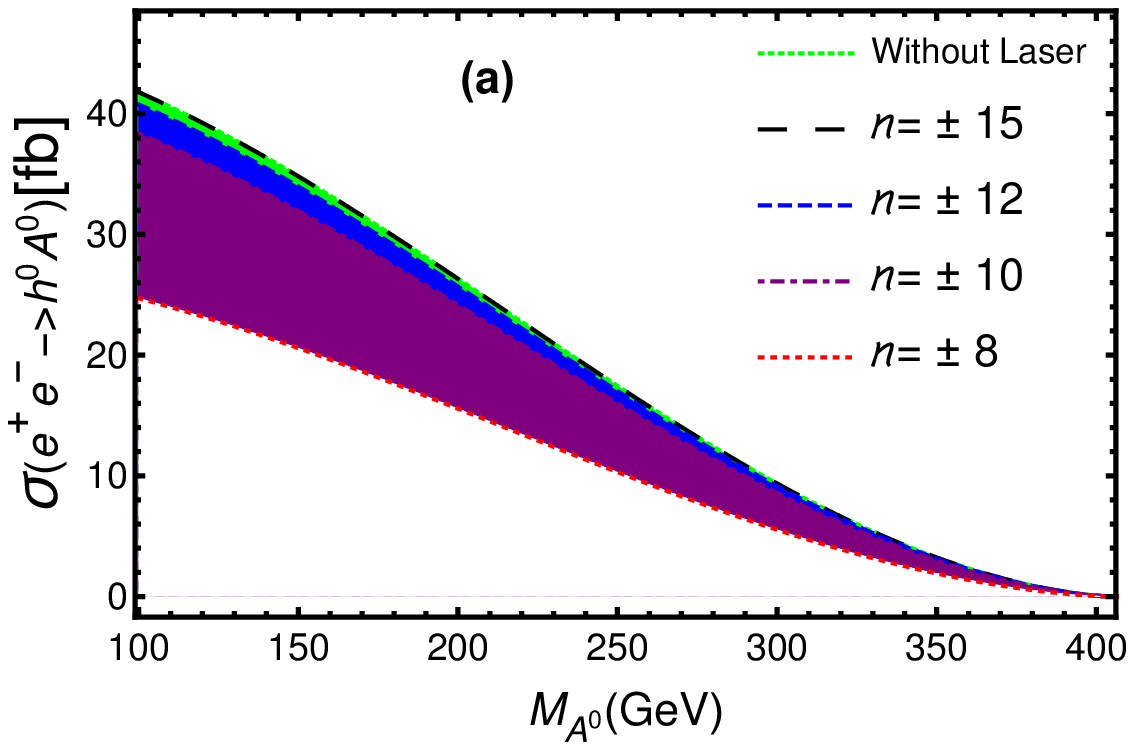}\hspace*{0.4cm}
      \includegraphics[scale=0.58]{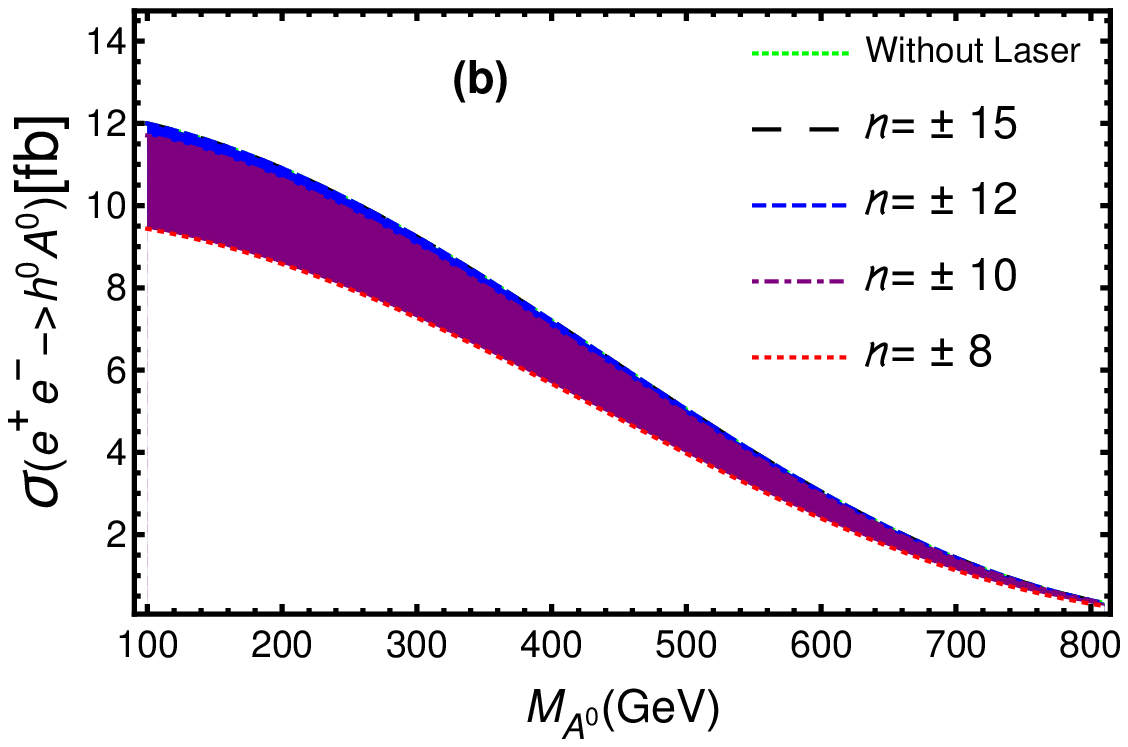}\par\vspace*{0.5cm}
      \includegraphics[scale=0.6]{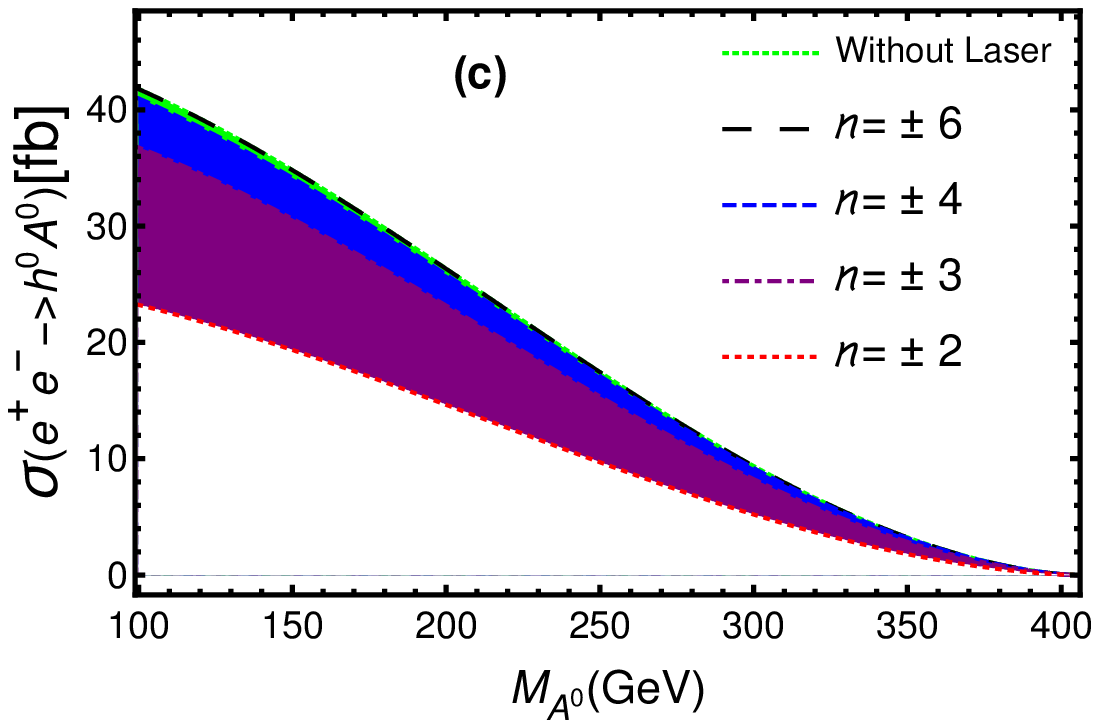}\hspace*{0.4cm}
      \includegraphics[scale=0.58]{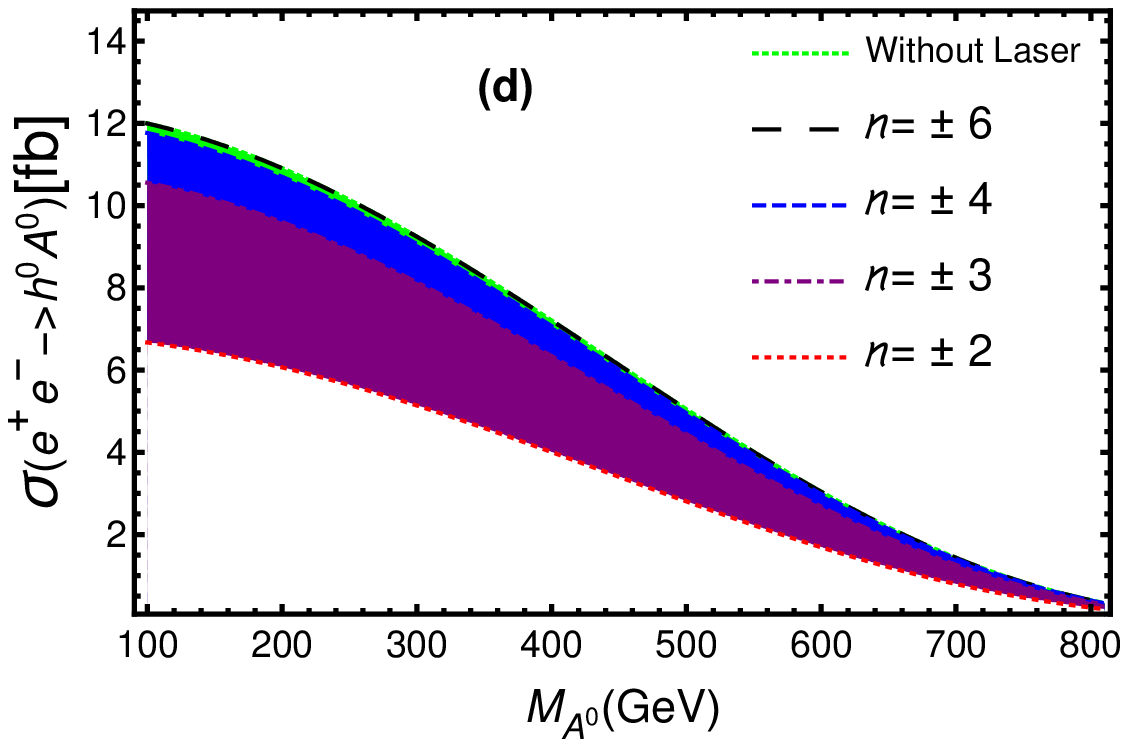}
        \caption{Variation of the laser-assisted total cross section of the process ${e}^{+}{e}^{-}\rightarrow h^{0}A^{0}$ as a function of the CP-odd Higgs-boson mass for different number of exchanged photons and for two typical centre of mass energies:  $500\,GeV$(left) and $1\,TeV$(right). The CP-even Higgs mass and the laser field strength are chosen as $M_{h^{0}}=95\,GeV$ and $\varepsilon_{0}=10^{5}\,V.cm^{-1}$, respectively. \textbf{Nd:YAG laser} $(\omega=1.17\, eV)$ is used in (a) and (b), while \textbf{He:Ne laser} $(\omega=2\, eV)$ is used in (c) and (d).}
        \label{fig5}
\end{figure}
Figure \ref{fig5} illustrates the behavior of the total cross section inside a circularly polarized electromagnetic field as a function of the CP-odd Higgs mass. The laser parameters used in this figure are similar to those used in the previous one. We remark that the cross section behaves in the same manner. However, its orders of magnitude are totally different. The maximum cross section reaches $\sigma_{max}=41.5\,[fb]$ and $\sigma_{max}=11.96\,[fb]$ for $\sqrt{s}=500\,GeV$ (left panel) and $\sqrt{s}=1\,TeV$ (right panel), respectively. Another important remark is that the total cross section drops rapidly in figure \ref{fig5}(a) and \ref{fig5}(c) as compared to the \ref{fig4}(a) and \ref{fig4}(c), respectively. For $\sqrt{s}=1\,TeV$, there is not much difference between figure \ref{fig4} and figure \ref{fig5} in terms of the masses which correspond to zero cross section. For instance, it becomes null whenever the CP-odd Higgs mass overcomes $M_{A^{0}}=400\,GeV$ and $M_{A^{0}}=800\,GeV$ for the left and right panel, respectively.
\subsection{The process ${e}^{+}{e}^{-}\rightarrow A^{0}H^{0}$}
\begin{figure}[H]
  \centering
      \includegraphics[scale=0.7]{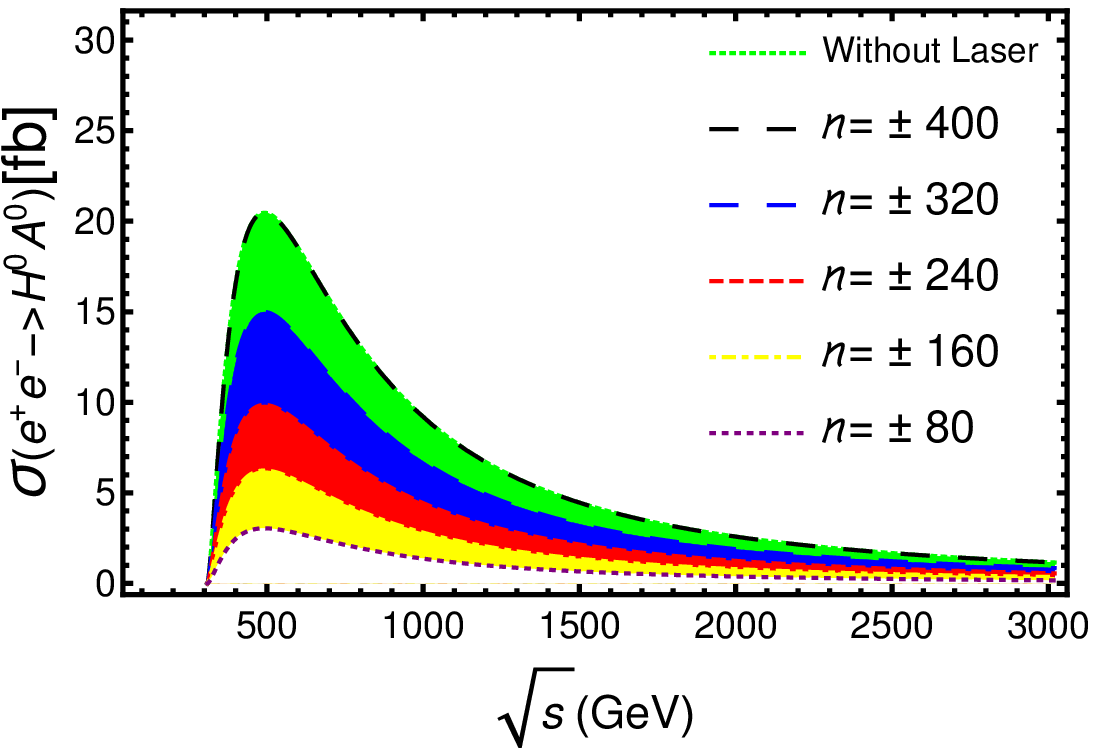}\hspace*{0.5cm}
      \includegraphics[scale=0.7]{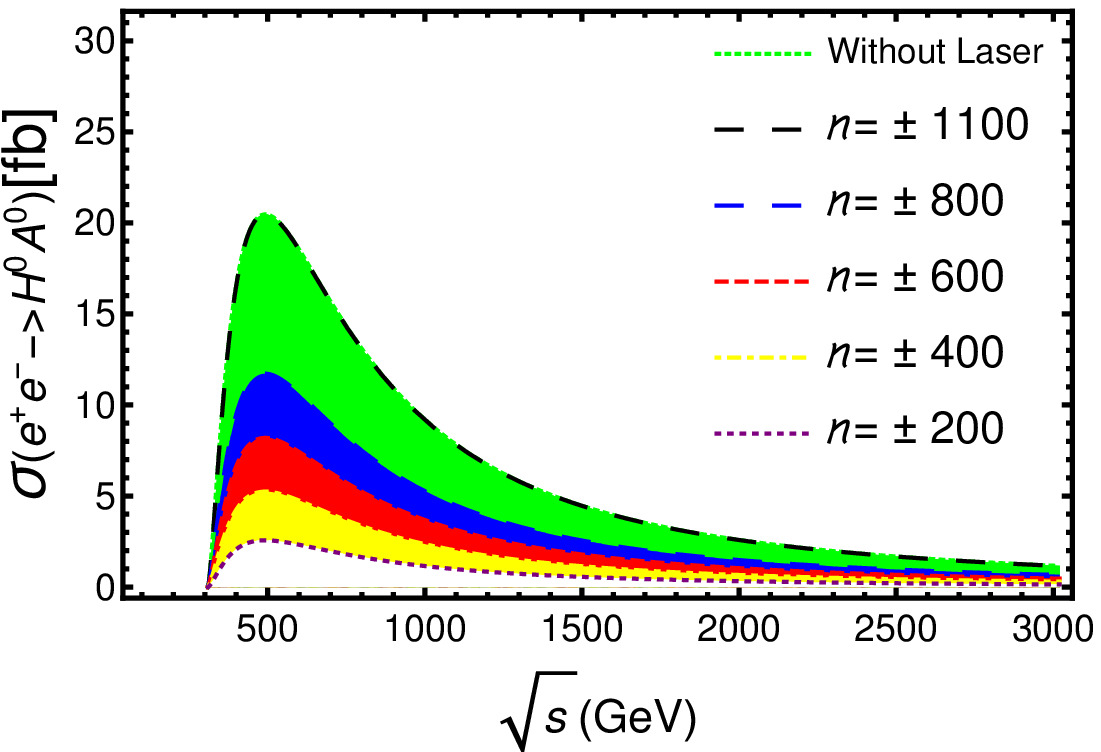}
        \caption{Variation of the laser-assisted total cross section of ${e}^{+}{e}^{-}\rightarrow H^{0}A^{0}$ as a function of the centre of mass energy for different exchanged photons number and by taking $M_{A^{0}}=92.20\,GeV$ and $M_{H^{0}}=212\,GeV$. The \textbf{He:Ne laser} $(\omega=2\, eV)$ with a strength $\varepsilon_{0}=10^{7}\,V.cm^{-1}$ is used in the left panel, while the \textbf{Nd:YAG laser} $(\omega=1.17\, eV)$ with  $\varepsilon_{0}=10^{7}\,V.cm^{-1}$ is used in the right panel.}
        \label{fig6}
\end{figure}

In this part, we will deal with the process $e^{+}e^{-}\rightarrow H^{0}A^{0}$ inside an electromagnetic field with circular polarization.
In this case, $h^{0}$ is considered as being the standard model-like Higgs boson with a mass $125\,GeV$, while $H^{0}$ is the CP-even Higgs boson.
Except in figures in which they are variables, the masses of the produced neutral Higgs bosons are chosen as $M_{A^{0}}=98.20\,GeV$ and $M_{H^{0}}=212\,GeV$. $\beta$ and $\alpha$ are chosen such that $\sin(\beta-\alpha)=0.99679$ \cite{Benbrik}.
We start our discussion by analyzing the effect of the laser field with different strengths and frequencies on the partial total cross section.
\small
\begin{table}[H]
\centering
 \caption{\label{tab1}Laser-assisted partial total cross section as a function of the number of exchanged photons for different laser field strengths and frequencies. The centre of mass energy and the neutral Higgs masses are chosen as: $\sqrt{s}=500\,GeV$ ; $M_{H^{0}}=212\,GeV$ ; $M_{A^{0}}=98.20\,GeV$.}
 \small
\begin{tabular}{ccccccccc}
\hline
  & & $ \sigma^{n}(e^{+}e^{-}\rightarrow H^{0}A^{0}) $[fb]  & &$ \sigma^{n}(e^{+}e^{-}\rightarrow H^{0}A^{0}) $[fb] & & $ \sigma^{n}(e^{+}e^{-}\rightarrow H^{0}A^{0}) $[fb] &  \\
Laser's type  & n &  $ \varepsilon_{0}=10^{5}\,V.cm^{-1} $ & n &$\varepsilon_{0}=10^{6}\,V.cm^{-1} $& n & $\varepsilon_{0}=10^{7}\,V.cm^{-1} $ & \\
 \hline
 &$\pm6$ & $ 0 $ & $\pm48$ & $ 0 $ & $\pm480$ & $ 0 $ \\
 &$\pm5$ & $ 0 $ & $\pm40$ & $ 0 $ & $\pm400$ & $ 0 $ \\
   He:Ne laser &$\pm4$ & $ 0.851739 $ & $\pm32$ & $ 0.874665 $ & $\pm320$ & $ 0.014779 $ \\
  $ \omega=2\, eV $ &$\pm3$ & $ 3.05466 $ & $\pm24$ & $ 0.0723476 $ & $\pm240$ & $ 0.000136283 $ \\
   &$\pm2$ & $ 4.30554 $ & $\pm16$ & $ 0.398907 $ & $\pm160$ & $ 0.0411635 $ \\
    &$\pm1$ & $ 0.390639 $ & $\pm8$ & $ 0.277157 $ & $\pm80$ & $ 0.0380871 $ \\
    &$0$ & $ 2.95024  $ & $0$ & $ 0.324287 $  & $0$ & $ 0.0340363 $ \\
     \hline
     &$\pm18$ & $ 0 $ & $\pm150$ & $ 0 $ & $\pm1300$ & $ 0 $ \\
 &$\pm15$ & $ 0 $ & $\pm125$ & $ 0 $ & $\pm1100$ & $ 0 $ \\
  $ CO_{2} $ laser  &$\pm12$ & $ 0.113047 $ & $\pm100$ & $ 0.364843 $ & $\pm800$ & $ 0.0147779 $ \\
  $ \omega=1.17\, eV $  &$\pm9$ & $ 1.87084 $ & $\pm75$ & $ 0.186565 $ & $\pm600$ & $ 0.0155667 $ \\
   &$\pm6$ & $ 0.081955 $ & $\pm50$ & $ 0.0677844 $ & $\pm400$ & $ 0.00279483 $ \\
   &$\pm3$ & $ 0.2412 $ & $\pm25$ & $ 0.071085 $ & $\pm200$ & $ 0.00257665 $ \\
    &$0$ & $ 1.27201  $ & $0$ & $ 0.0497559 $  & $0$ & $ 0.0114089 $ \\
     \hline
\end{tabular}
\end{table}
\normalsize
Table \ref{tab1} represents the dependence of the partial total cross section on the number of transferred photons for different laser field strengths and frequencies by taking the centre of mass energy as $\sqrt{s}=500\,GeV$. As illustrated in the previous section, the partial total cross section presents cutoffs at two symmetric edges. This means that the cross section which corresponds to emission of photons has the same values as that corresponds to the absorption. These cutoffs are similar to those obtained in section (3). However, the orders of magnitude of the partial total cross section are different.
In addition, the cutoff number increases either by decreasing the frequency or by increasing the laser field strength. For instance, for $ \varepsilon_{0}=10^{5}\,V.cm^{-1} $, the cutoffs are $\pm 400$ and $\pm1100$ for the \textbf{ He:Ne laser} and \textbf{$ CO_{2} $ laser}, respectively. The behavior of the summed laser-assisted total cross section as a function of the collider's centre of mass energy is illustrated in figure \ref{fig6}.
According to figure \ref{fig6}, the cross section behaves in the same manner as in the case ${e}^{+}{e}^{-}\rightarrow h^{0}A^{0}$ where $H^{0}$ is considered as the SM Higgs boson. Moreover, it begins to increase immediately when the centre of mass energy overcomes a threshold value which is around $\sqrt{s}=300\,GeV$. It reaches its maximum at $\sqrt{s}=500\,GeV$, and due to the phase space suppression, it decreases progressively as long as $\sqrt{s}$ raises.
It is clear that the \textbf{Nd:YAG laser} with a strength $\varepsilon_{0}=10^{7}\,V.cm^{-1}$ strongly affects the total cross section, and a big number of photons, which is approximately $\pm 1100$, is required to fulfill the sum-rule. In contrast, for the \textbf{He:Ne laser}, it requires just $\pm400$ photons to reach the laser-free cross section.
Let's focus now on the dependence of the laser-assisted total cross section on the masses of the produced neutral Higgs bosons.
\begin{figure}[H]
  \centering
      \includegraphics[scale=0.55]{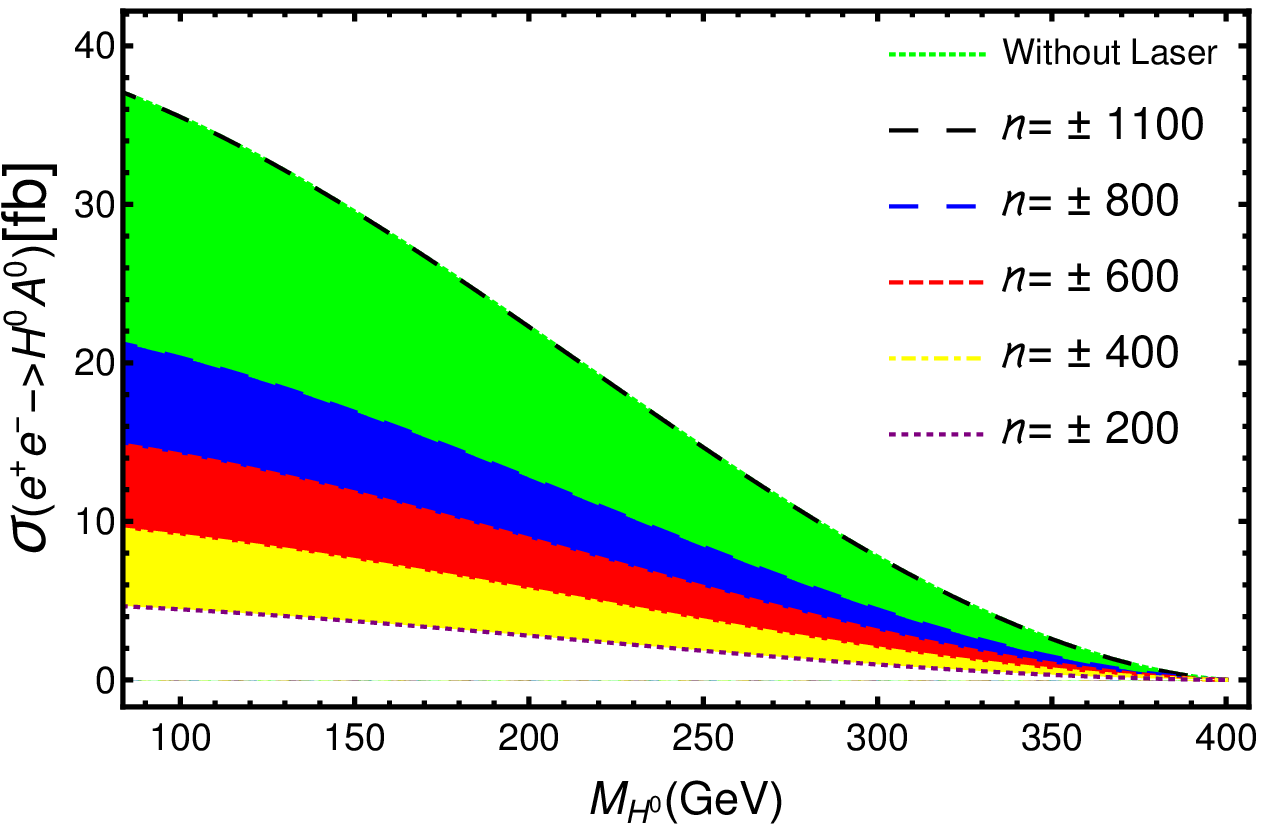}\hspace*{0.5cm}
      \includegraphics[scale=0.58]{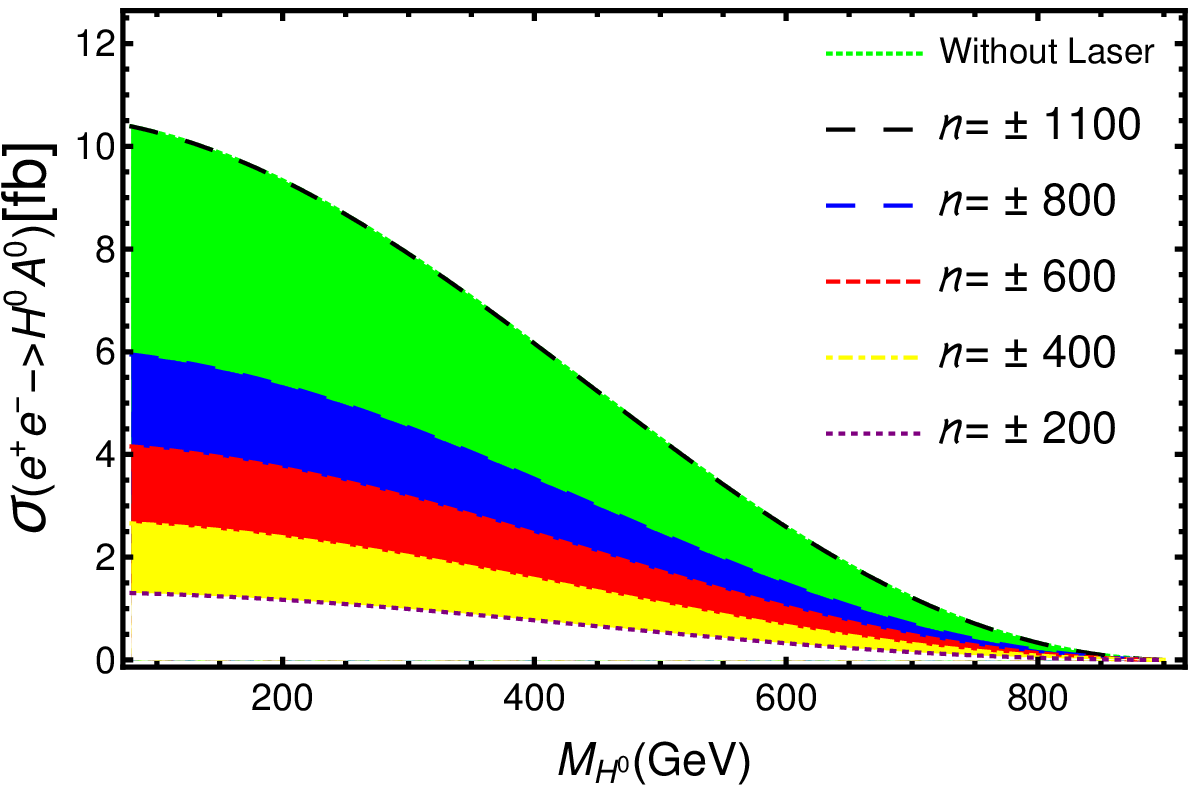}
        \caption{Laser-assisted total cross section of ${e}^{+}{e}^{-}\rightarrow H^{0}A^{0}$ versus the CP-even Higgs-boson mass for different number of exchanged photons by taking $M_{A^{0}}=98.20\,GeV$ and for two centre of mass energies:  $500\,GeV$ (left panel) and $1\,TeV$ (right panel). The laser field strength and its frequency are taken as: $\varepsilon_{0}=10^{7}\,V.cm^{-1}$ and $\omega=1.17\, eV$ (Nd:YAG laser).}
        \label{fig7}
\end{figure}
\begin{figure}[H]
  \centering
      \includegraphics[scale=0.55]{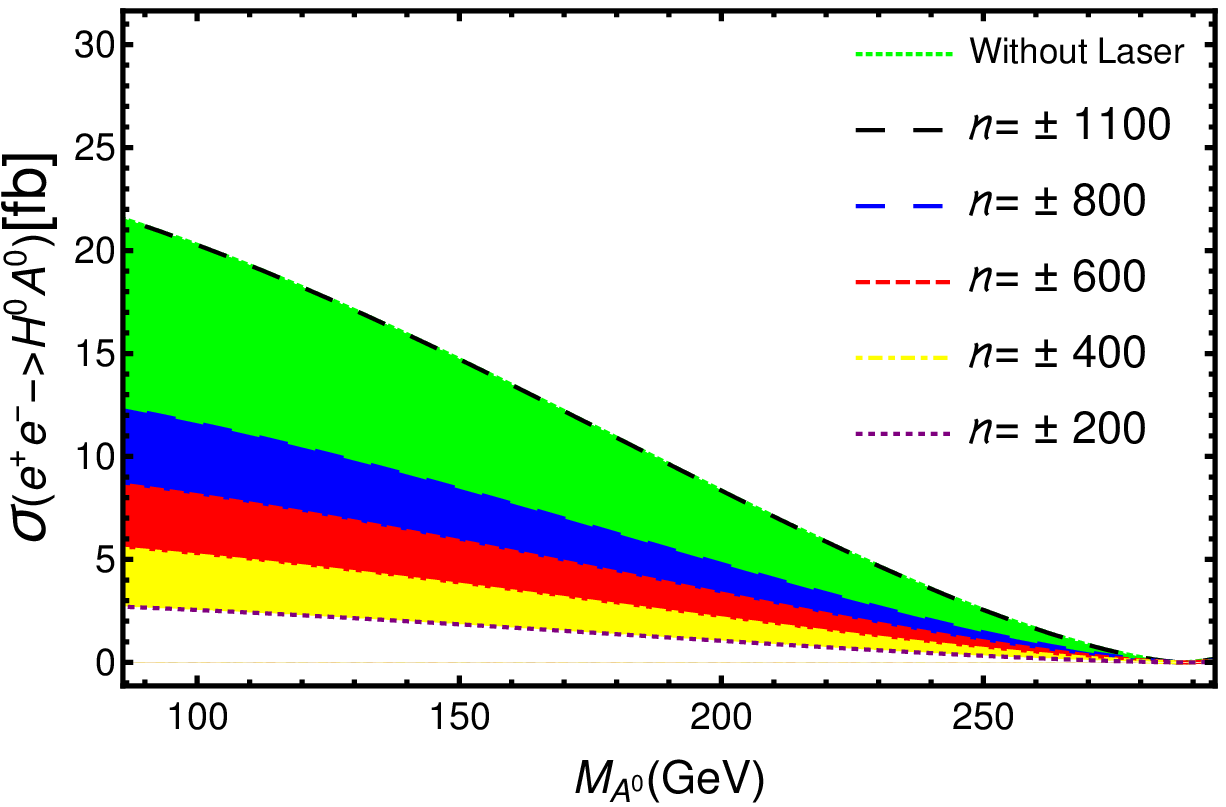}\hspace*{0.5cm}
      \includegraphics[scale=0.55]{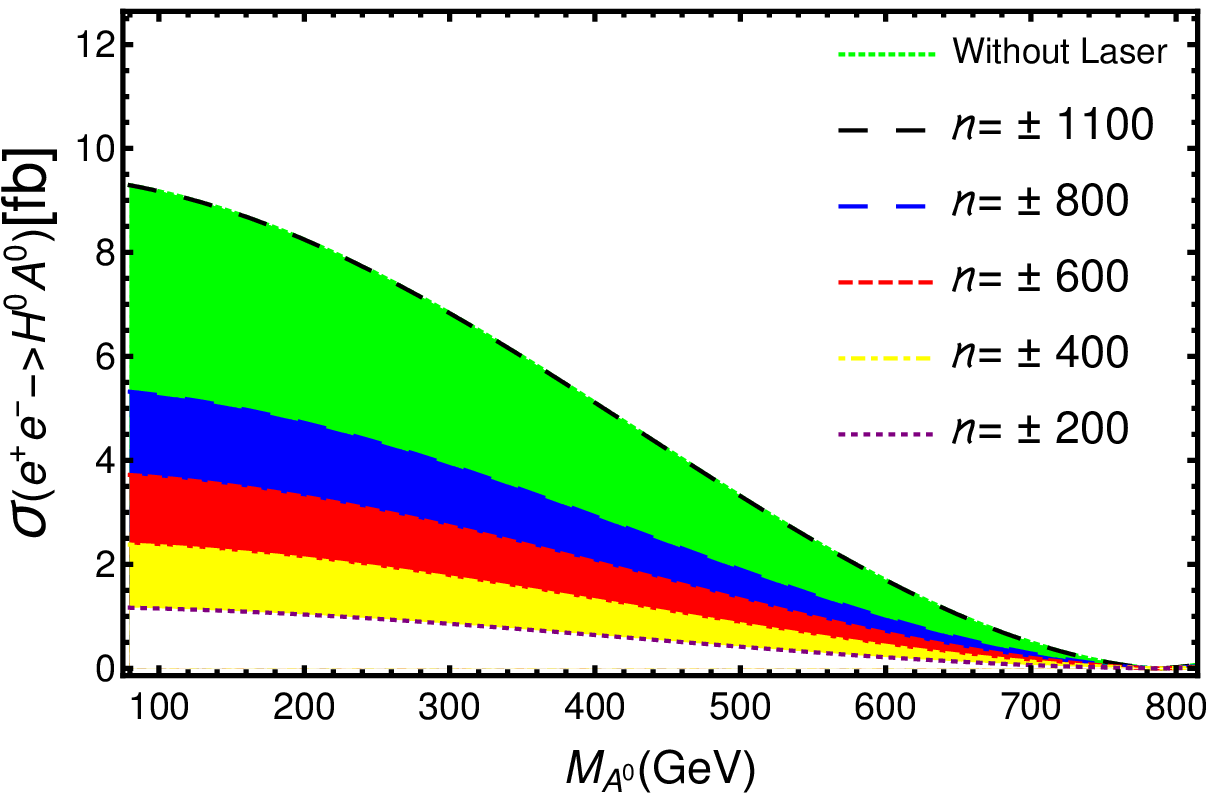}
        \caption{Dependence of the laser-assisted total cross section of the process ${e}^{+}{e}^{-}\rightarrow H^{0}A^{0}$ on the CP-odd Higgs-boson mass for different number of exchanged photons by taking $M_{H^{0}}=212\,GeV$ and for two centre of mass energies: $500\,GeV$ (left panel) and $1\,TeV$ (right panel). The laser field strength and its frequency are taken as: $\varepsilon_{0}=10^{7}\,V.cm^{-1}$ and $\omega=1.17\, eV$ (Nd:YAG laser).}
        \label{fig8}
\end{figure}
Figures \ref{fig7} and \ref{fig8} successively represent the effect of the \textbf{Nd:YAG laser} with a strength $\varepsilon_{0}=10^{7}\,V.cm^{-1}$ on the total cross section as a function of $M_{H^{0}}$ and $M_{A^{0}}$.
Regardless of the laser field parameters and the centre of mass energy, the total cross section is maximal at $M_{H^{0}}=100\,GeV$ and $M_{A^{0}}=100\,GeV$ in figure \ref{fig7} and figure \ref{fig8}, respectively. Then, it decreases progressively a much as the produced neutral Higgs mass increases, and it becomes null at $M_{H^{0}}=380\,GeV$ and $M_{A^{0}}=280\,GeV$ for $\sqrt{s}=500\,GeV$.
In addition, the order of magnitude of the total cross section depends on the centre of mass energy of the collider. Moreover, this order is higher for $\sqrt{s}=500\,GeV$ as compared to that where $\sqrt{s}=1\,TeV$. Furthermore, the maximum number of photons to be exchanged in order to reach the well known sum-rule does not depend on $\sqrt{s}$, and it only depends on the laser field strength and its frequency. For instance, for $\sqrt{s}=500\,GeV$, the maximum cross section which corresponds to $\pm 1100$ is successively $\sigma_{max}=35.92\,GeV$ and $\sigma_{max}=20.61\,GeV$ in figure \ref{fig7} (left panel) and figure \ref{fig8} (left panel).
\begin{figure}[H]
  \centering
      \includegraphics[scale=0.50]{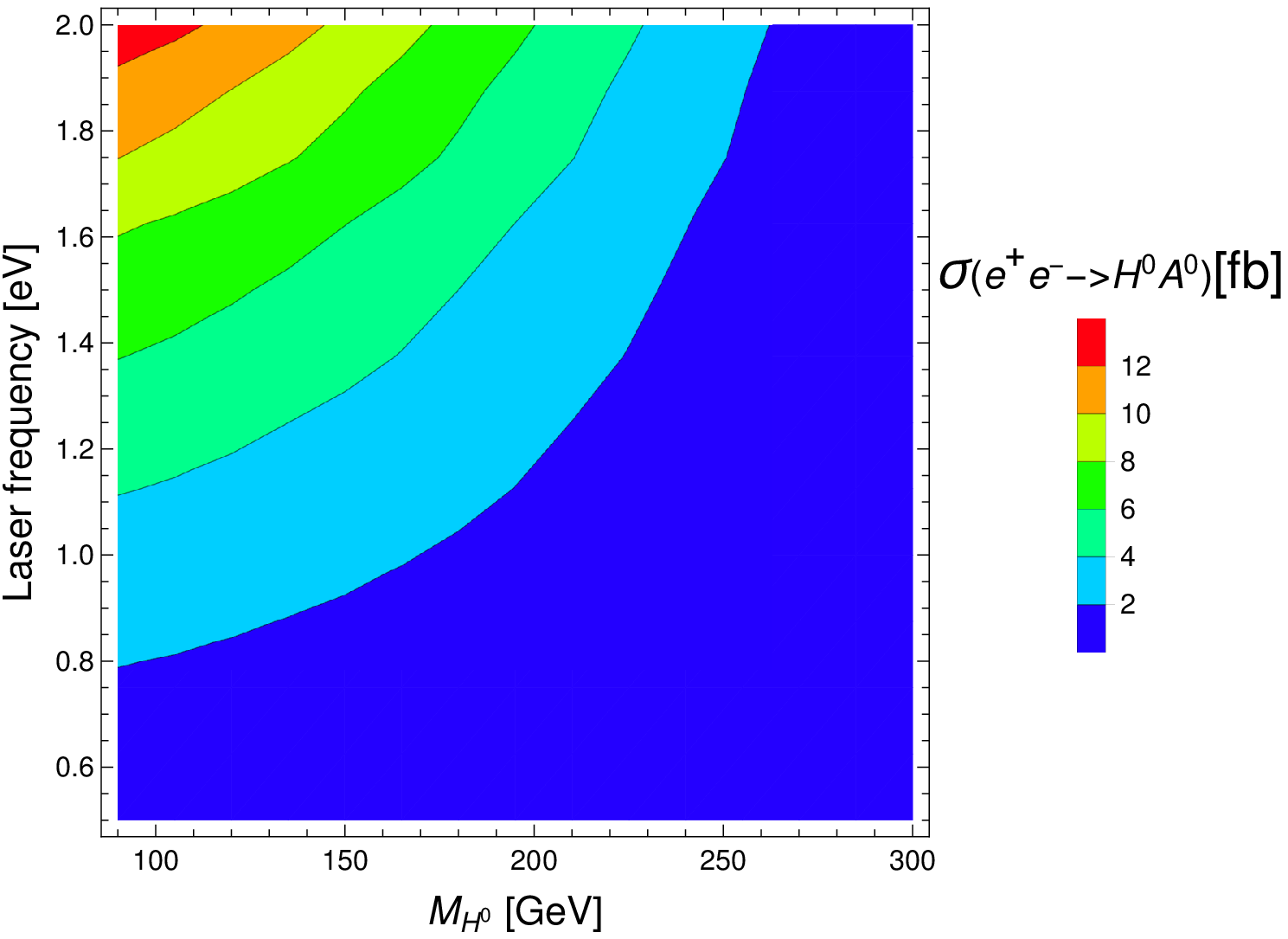}\hspace*{0.5cm}
      \includegraphics[scale=0.50]{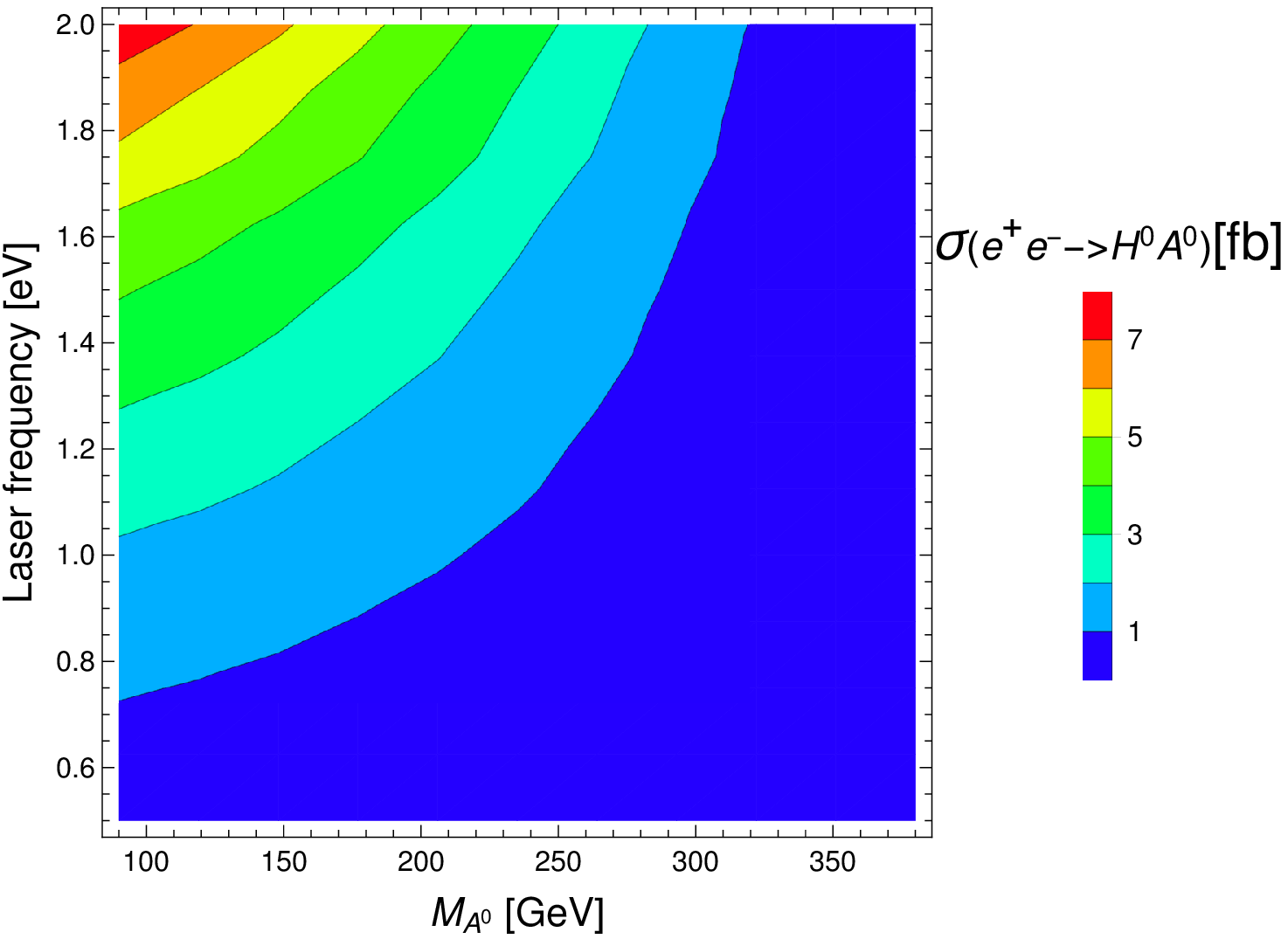}
        \caption{Behavior of the laser-assisted total cross section as a function of CP-even (left panel) or CP-odd mass (right panel) and the laser field frequency by taking the number of transferred photons as $n=\pm\,20$, the centre of mass energy as $\sqrt{s}= 500\, GeV$, the laser field strength as $\varepsilon_{0}=10^{6}\,V.cm^{-1}$. (Left): $ M_{A^{0}}=98.20\, GeV $; (Right): $ M_{H^{0}}=212\, GeV $.}
        \label{fig9}
\end{figure}
\begin{figure}[H]
  \centering
      \includegraphics[scale=0.47]{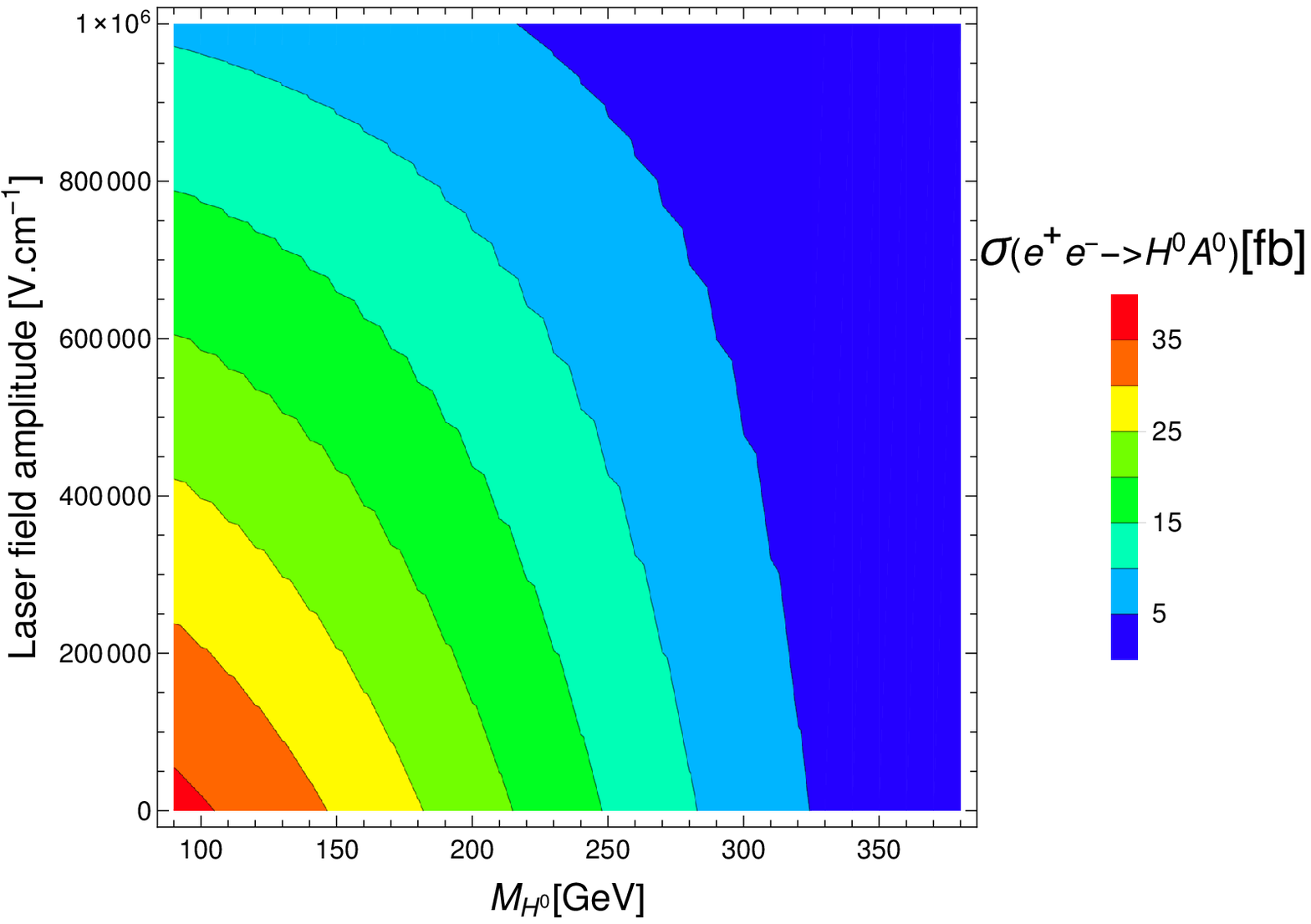}\hspace*{0.5cm}
      \includegraphics[scale=0.44]{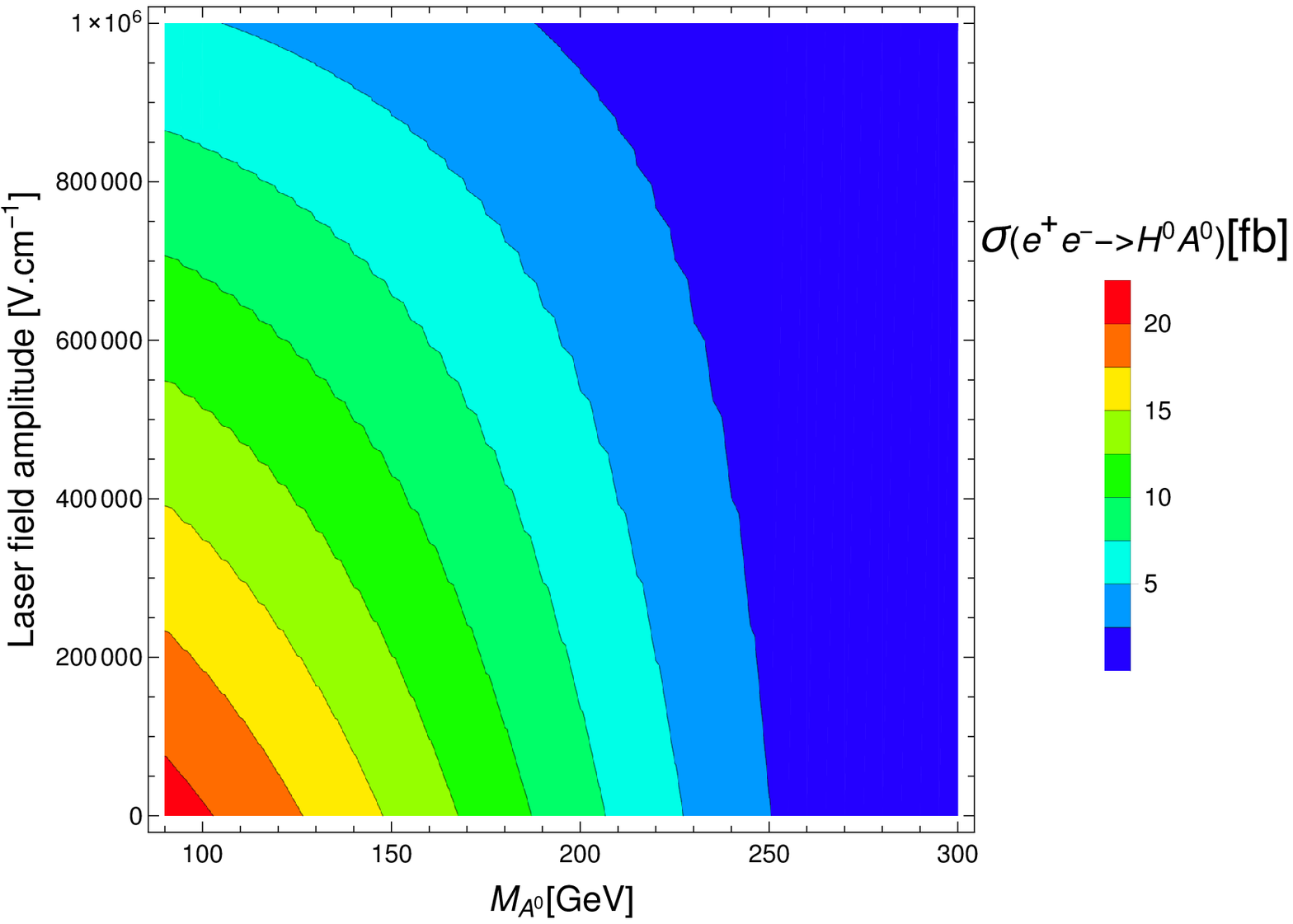}
        \caption{Behavior of the laser-assisted total cross section as a function of CP-even (left panel) or CP-odd mass (right panel) and the laser field amplitude by taking the number of transferred photons as $n=\pm\,20$, the centre of mass energy as $\sqrt{s}= 500\, GeV$ and the laser field frequency as $\omega=1.17\,eV$. (left panel): $ M_{A^{0}}=98.20\, GeV $; (right panel): $ M_{H^{0}}=212\, GeV $.}
        \label{fig10}
\end{figure}
Figures \ref{fig9} and \ref{fig10} represent the simultaneous dependence of the laser-assisted total cross section on the mass of each one of the produced neutral Higgs bosons and on the laser field frequency or its strength by taking the centre of mass energy as $\sqrt{s}= 500\, GeV$. The red zone corresponds to the maximum cross section, while the blue one corresponds to its minimum. In addition, we observe that, for a given neutral Higgs mass, the cross section decreases by whether increasing the laser field strength or by decreasing its frequency.
\section{Conclusion}
In this work, we have investigated the pair production process of neutral Higgs boson in THDM inside an electromagnetic field with circular polarization. We have made an exact and efficient analytical calculation of the differential cross section. For the computation of the total cross section, two benchmark points are considered for the THDM parameters. We have found that, regardless of the laser parameters, neutral Higgs masses and centre of mass energy, the partial total cross section is bounded by two symmetric cutoffs. In addition, the summed total cross section is higher for $\sqrt{s}=500\,GeV$ as compared to $\sqrt{s}=1\,TeV$ regardless of the chosen benchmark points.
Moreover, it decreases by several orders of magnitude, and this decreasing process depends on the number of exchanged photons, laser field strength and its frequency.
\section{Appendix}
\tiny
\begin{eqnarray}
\Delta_{2}&=&\nonumber\dfrac{e^2 }{(k.p_{1})(k.p_{2})}\Big[
    (4 (k.p_{1}) (k.p_{2}) ((a_{2}.k_{1})^2 g_{a}^{e^{2}} (k.p_{1}) (k.p_{2}) - 2 (a_{2}.k_{1}) (a_{2}.k_{2}) g_{a}^{e^{2}} (k.p_{1}) (k.p_{2}) +
        (a_{2}.k_{2})^2 g_{a}^{e^{2}}  (k.p_{1}) (k.p_{2}) + (a_{2}.k_{1})^2 g_{v}^{e^{2}} (k.p_{1}) (k.p_{2}) \\&- &\nonumber
        2 (a_{2}.k_{1}) (a_{2}.k_{2}) g_{v}^{e^{2}} (k.p_{1}) (k.p_{2}) + (a_{2}.k_{2})^2 g_{v}^{e^{2}} (k.p_{1}) (k.p_{2})
        (a_{1}.k_{1})^2 (g_{a}^{e^{2}} + g_{v}^{e^{2}}) (k.p_{1}) (k.p_{2}) +
        (a_{1}.k_{2})^2 (g_{a}^{e^{2}} + g_{v}^{e^{2}}) (k.p_{1}) (k.p_{2}) + (a_{1}.p_{1})\\&\times &\nonumber (a_{1}.p_{2}) g_{a}^{e^{2}} (k.k_{1})^2
        (a_{1}.p_{1}) (a_{1}.p_{2}) g_{v}^{e^{2}} (k.k_{1})^2 -
        (a_{1}.k_{1}) (g_{a}^{e^{2}} +
           g_{v}^{e^{2}}) (2 (a_{1}.k_{2}) (k.p_{1}) (k.p_{2}) + ((a_{1}.p_{2}) (k.p_{1}) + (a_{1}.p_{1})  (k.p_{2})) ((k.k_{1}) -
              (k.k_{2}))) \\&+ &\nonumber
        (a_{1}.k_{2}) (g_{a}^{e^{2}} + g_{v}^{e^{2}}) ((a_{1}.p_{2}) (k.p_{1}) + (a_{1}.p_{1}) (k.p_{2})) ((k.k_{1}) - (k.k_{2}))
        2 (a_{1}.p_{1}) (a_{1}.p_{2}) g_{a}^{e^{2}} (k.k_{1}) (k.k_{2}) - 2 (a_{1}.p_{1}) (a_{1}.p_{2}) g_{v}^{e^{2}} (k.k_{1}) (k.k_{2}) +
        (a_{1}.p_{1})\\&\times &\nonumber (a_{1}.p_{2}) g_{a}^{e^{2}} (k.k_{2})^2   (a_{1}.p_{1}) (a_{1}.p_{2}) g_{v}^{e^{2}} (k.k_{2})^2 -
        a^2 g_{a}^{e^{2}} (k.p_{1})^2 M_{A^{0}}^{2} - a^2 g_{v}^{e^{2}} (k.p_{1})^2 M_{A^{0}}^{2} -
        2 a^2 g_{a}^{e^{2}} (k.p_{1}) (k.p_{2}) M_{A^{0}}^{2}  2 a^2 g_{v}^{e^{2}} (k.p_{1}) (k.p_{2}) M_{A^{0}}^{2} -
        a^2 \\&\times &\nonumber g_{a}^{e^{2}} (k.p_{2})^2 M_{A^{0}}^{2} - a^2 g_{v}^{e^{2}} (k.p_{2})^2 M_{A^{0}}^{2} +
        a^2 g_{a}^{e^{2}} (k.k_{1})^2 me^2 - a^2 g_{v}^{e^{2}} (k.k_{1})^2  me^2 -
        2 a^2 g_{a}^{e^{2}} (k.k_{1}) (k.k_{2}) me^2 + 2 a^2 g_{v}^{e^{2}} (k.k_{1}) (k.k_{2}) me^2 +
        a^2 g_{a}^{e^{2}}\\&\times &\nonumber (k.k_{2})^2 me^2 - a^2 g_{v}^{e^{2}} (k.k_{2})^2 me^2
        a^2 g_{a}^{e^{2}} (k.k_{1})^2 (p_{1}.p_{2}) - a^2 g_{v}^{e^{2}} (k.k_{1})^2 (p_{1}.p_{2}) +
        2 a^2 g_{a}^{e^{2}} (k.k_{1}) (k.k_{2}) (p_{1}.p_{2}) + 2 a^2 g_{v}^{e^{2}} (k.k_{1}) (k.k_{2}) (p_{1}.p_{2})-
        a^2 g_{a}^{e^{2}} \\&\times &\nonumber (k.k_{2})^2 (p_{1}.p_{2}) - a^2 g_{v}^{e^{2}} (k.k_{2})^2 (p_{1}.p_{2}) +
        a^2 g_{a}^{e^{2}} (k.p_{1}) (k.k_{1}) (p_{1}.k_{1}) + a^2 g_{v}^{e^{2}} (k.p_{1}) (k.k_{1}) (p_{1}.k_{1}) +
        a^2 g_{a}^{e^{2}} (k.p_{2}) (k.k_{1}) (p_{1}.k_{1}) + a^2 g_{v}^{e^{2}} (k.p_{2}) (k.k_{1})\\&\times &\nonumber (p_{1}.k_{1}) -
        a^2 g_{a}^{e^{2}} (k.p_{1}) (k.k_{2}) (p_{1}.k_{1}) - a^2 g_{v}^{e^{2}} (k.p_{1})  (k.k_{2}) (p_{1}.k_{1}) -
        a^2 g_{a}^{e^{2}} (k.p_{2}) (k.k_{2}) (p_{1}.k_{1}) - a^2 g_{v}^{e^{2}} (k.p_{2}) (k.k_{2}) (p_{1}.k_{1}) -
        a^2 g_{a}^{e^{2}} (k.p_{1}) (k.k_{1})\\&\times &\nonumber (p_{1}.k_{2}) - a^2 g_{v}^{e^{2}} (k.p_{1}) (k.k_{1}) (p_{1}.k_{2}) -
        a^2 g_{a}^{e^{2}} (k.p_{2}) (k.k_{1}) (p_{1}.k_{2}) - a^2 g_{v}^{e^{2}} (k.p_{2}) (k.k_{1}) (p_{1}.k_{2}) +
        a^2 g_{a}^{e^{2}} (k.p_{1})  (k.k_{2}) (p_{1}.k_{2}) + a^2 g_{v}^{e^{2}} (k.p_{1}) (k.k_{2})\\&\times &\nonumber (p_{1}.k_{2}) +
        a^2 g_{a}^{e^{2}} (k.p_{2}) (k.k_{2})(p_{1}.k_{2}) + a^2 g_{v}^{e^{2}} (k.p_{2}) (k.k_{2}) (p_{1}.k_{2}) +
        a^2 g_{a}^{e^{2}} (k.p_{1}) (k.k_{1}) (p_{2}.k_{1}) + a^2 g_{v}^{e^{2}} (k.p_{1}) (k.k_{1}) (p_{2}.k_{1}) +
        a^2 g_{a}^{e^{2}} (k.p_{2}) (k.k_{1})\\&\times &\nonumber (p_{2}.k_{1}) + a^2 g_{v}^{e^{2}} (k.p_{2}) (k.k_{1}) (p_{2}.k_{1}) -
        a^2 g_{a}^{e^{2}} (k.p_{1}) (k.k_{2}) (p_{2}.k_{1}) - a^2 g_{v}^{e^{2}} (k.p_{1}) (k.k_{2}) (p_{2}.k_{1}) -
        a^2 g_{a}^{e^{2}} (k.p_{2}) (k.k_{2}) (p_{2}.k_{1}) - a^2 g_{v}^{e^{2}}  (k.p_{2}) (k.k_{2})\\&\times &\nonumber (p_{2}.k_{1}) -
        a^2 g_{a}^{e^{2}} (k.p_{1}) (k.k_{1}) (p_{2}.k_{2}) - a^2 g_{v}^{e^{2}} (k.p_{1}) (k.k_{1}) (p_{2}.k_{2}) -
        a^2 g_{a}^{e^{2}} (k.p_{2}) (k.k_{1})  (p_{2}.k_{2}) - a^2 g_{v}^{e^{2}} (k.p_{2}) (k.k_{1}) (p_{2}.k_{2}) +
        a^2 g_{a}^{e^{2}} (k.p_{1}) (k.k_{2})\\&\times &\nonumber (p_{2}.k_{2}) + a^2 g_{v}^{e^{2}} (k.p_{1}) (k.k_{2}) (p_{2}.k_{2}) +
        a^2 g_{a}^{e^{2}}  (k.p_{2}) (k.k_{2}) (p_{2}.k_{2}) + a^2 g_{v}^{e^{2}} (k.p_{2}) (k.k_{2}) (p_{2}.k_{2}) +
        a^2 (g_{a}^{e^{2}} + g_{v}^{e^{2}}) ((k.p_{1}) + (k.p_{2}))^2 (k_{1}.k_{2})) +
     g_{a}^{e} g_{v}^{e}\\&\times &\nonumber (k.p_{2}) ((k.p_{2}) (-3 ((k.k_{1}) - (k.k_{2})) ((p_{2}.k_{1}) - (p_{2}.k_{2})) +
           4 (k.p_{2}) (M_{A^{0}}^{2} - (k_{1}.k_{2}))) +
        (k.p_{1}) (-((k.k_{1}) - (k.k_{2})) ((p_{2}.k_{1}) - (p_{2}.k_{2})) + 8 (k.p_{2})\\&\times &\nonumber (M_{A^{0}}^{2} - (k_{1}.k_{2}))) +
        4 (k.p_{1})^2 (M_{A^{0}}^{2} - (k_{1}.k_{2}))) \epsilon(a_{1}, a_{2},
       k, p_{1}) +
     g_{a}^{e} g_{v}^{e} (k.p_{1})  (4 (k.p_{1})^2 (-M_{A^{0}}^{2} + (k_{1}.k_{2})) +
        (k.p_{2}) (((k.k_{1}) - (k.k_{2})) ((p_{1}.k_{1}) \\&- &\nonumber(p_{1}.k_{2})) + 4 (k.p_{2}) (-M_{A^{0}}^{2} + (k_{1}.k_{2})))
        (k.p_{1}) (3 ((k.k_{1}) - (k.k_{2})) ((p_{1}.k_{1}) - (p_{1}.k_{2})) + 8 (k.p_{2}) (-M_{A^{0}}^{2} + (k_{1}.k_{2})))) \epsilon(
       a_{1}, a_{2}, k, p_{2}) +
     g_{a}^{e} g_{v}^{e} (((k.p_{1}) \\&- &\nonumber (k.p_{2})) ((k.p_{1}) + (k.p_{2})) ((k.k_{1}) - (k.k_{2})) (p_{1}.p_{2}) \epsilon(
          a_{1}, a_{2}, k,
          k_{1}) - ((k.p_{1}) - (k.p_{2})) ((k.p_{1}) + (k.p_{2})) ((k.k_{1}) -
           (k.k_{2})) (p_{1}.p_{2}) \epsilon(a_{1}, a_{2}, k,
          k_{2}) \\&+ &\nonumber
        3 (k.p_{1})^2 (k.k_{1})^2 \epsilon(a_{1}, a_{2}, p_{1},
          p_{2}) -
        6 (k.p_{1}) (k.p_{2})  (k.k_{1})^2 \epsilon(a_{1}, a_{2}, p_{1},
          p_{2}) +
        3 (k.p_{2})^2 (k.k_{1})^2 \epsilon(a_{1}, a_{2}, p_{1},
          p_{2}) -
        6 (k.p_{1})^2 (k.k_{1}) (k.k_{2}) \epsilon(a_{1}, a_{2}, p_{1},
          p_{2}) \\&+ &\nonumber
        12 (k.p_{1}) (k.p_{2}) (k.k_{1}) (k.k_{2}) \epsilon(a_{1}, a_{2},
          p_{1}, p_{2}) -
        6 (k.p_{2})^2 (k.k_{1}) (k.k_{2}) \epsilon(a_{1}, a_{2}, p_{1},
          p_{2}) +
        3 (k.p_{1})^2 (k.k_{2})^2  \epsilon(a_{1}, a_{2}, p_{1},
          p_{2}) -
        6 (k.p_{1}) (k.p_{2}) (k.k_{2})^2\\&\times &\nonumber \epsilon(a_{1}, a_{2}, p_{1},
          p_{2}) +
        3 (k.p_{2})^2 (k.k_{2})^2 \epsilon(a_{1}, a_{2}, p_{1},
          p_{2}) +
        (k.p_{1})^2 (k.p_{2}) (k.k_{1}) \epsilon(a_{1}, a_{2}, p_{1},
          k_{1}) +
        5 (k.p_{1}) (k.p_{2})^2 (k.k_{1}) \epsilon(a_{1}, a_{2}, p_{1},
          k_{1}) -
        2 (k.p_{2})^3 (k.k_{1})\\&\times &\nonumber \epsilon(a_{1}, a_{2}, p_{1},
          k_{1}) -
        (k.p_{1})^2 (k.p_{2}) (k.k_{2}) \epsilon(a_{1}, a_{2}, p_{1},
          k_{1}) -
        5 (k.p_{1}) (k.p_{2})^2 (k.k_{2}) \epsilon(a_{1}, a_{2}, p_{1},
          k_{1}) +
        2 (k.p_{2})^3 (k.k_{2})  \epsilon(a_{1}, a_{2}, p_{1},
          k_{1}) -
        (k.p_{1})^2 (k.p_{2})\\&\times &\nonumber (k.k_{1}) \epsilon(a_{1}, a_{2}, p_{1},
          k_{2}) -
        5 (k.p_{1}) (k.p_{2})^2 (k.k_{1}) \epsilon(a_{1}, a_{2}, p_{1},
          k_{2}) +
        2 (k.p_{2})^3 (k.k_{1}) \epsilon(a_{1}, a_{2}, p_{1},
          k_{2}) +
        (k.p_{1})^2 (k.p_{2}) (k.k_{2}) \epsilon(a_{1}, a_{2}, p_{1},
          k_{2}) +
        5 (k.p_{1})\\&\times &\nonumber (k.p_{2})^2 (k.k_{2})  \epsilon(a_{1}, a_{2}, p_{1},
          k_{2}) -
        2 (k.p_{2})^3 (k.k_{2}) \epsilon(a_{1}, a_{2}, p_{1},
          k_{2}) +
        2 (k.p_{1})^3 (k.k_{1}) \epsilon(a_{1}, a_{2}, p_{2},
          k_{1}) -
        5 (k.p_{1})^2 (k.p_{2})  (k.k_{1}) \epsilon(a_{1}, a_{2}, p_{2},
          k_{1}) -
        (k.p_{1}) \\&\times &\nonumber(k.p_{2})^2 (k.k_{1}) \epsilon(a_{1}, a_{2}, p_{2},
          k_{1}) -
        2 (k.p_{1})^3 (k.k_{2}) \epsilon(a_{1}, a_{2}, p_{2},
          k_{1}) +
        5 (k.p_{1})^2  (k.p_{2}) (k.k_{2}) \epsilon(a_{1}, a_{2}, p_{2},
          k_{1}) +
        (k.p_{1}) (k.p_{2})^2 (k.k_{2}) \epsilon(a_{1}, a_{2}, p_{2},
          k_{1}) \\&- &\nonumber
        2 (k.p_{1})^3 (k.k_{1}) \epsilon(a_{1}, a_{2}, p_{2},
          k_{2}) +
        5 (k.p_{1})^2 (k.p_{2}) (k.k_{1}) \epsilon(a_{1}, a_{2}, p_{2},
          k_{2}) +
        (k.p_{1}) (k.p_{2})^2 (k.k_{1}) \epsilon(a_{1}, a_{2}, p_{2},
          k_{2}) +
        2 (k.p_{1})^3 (k.k_{2}) \epsilon(a_{1}, a_{2}, p_{2},
          k_{2}) \\&- &\nonumber
        5 (k.p_{1})^2 (k.p_{2}) (k.k_{2}) \epsilon(a_{1}, a_{2}, p_{2},
          k_{2}) -
        (k.p_{1}) (k.p_{2})^2 (k.k_{2}) \epsilon(a_{1}, a_{2}, p_{2},
          k_{2}) +
        4 (a_{2}.k_{1}) (k.p_{1}) (k.p_{2}) (k.k_{1}) \epsilon(a_{1}, k,
          p_{1}, p_{2}) -
        4 (a_{2}.k_{2}) (k.p_{1}) (k.p_{2})\\&\times &\nonumber (k.k_{1}) \epsilon(a_{1}, k,
          p_{1}, p_{2}) -
        4 (a_{2}.k_{1}) (k.p_{1}) (k.p_{2}) (k.k_{2}) \epsilon(a_{1}, k,
          p_{1}, p_{2}) +
        4 (a_{2}.k_{2}) (k.p_{1}) (k.p_{2}) (k.k_{2}) \epsilon(a_{1}, k,
          p_{1}, p_{2}) -
        4 (a_{2}.k_{1}) (k.p_{1})^2 (k.p_{2})\\&\times &\nonumber  \epsilon(a_{1}, k, p_{1},
          k_{1}) +
        4 (a_{2}.k_{2}) (k.p_{1})^2 (k.p_{2}) \epsilon(a_{1}, k, p_{1},
          k_{1}) -
        4 (a_{2}.k_{1}) (k.p_{1}) (k.p_{2})^2 \epsilon(a_{1}, k, p_{1},
          k_{1}) +
        4 (a_{2}.k_{2}) (k.p_{1}) (k.p_{2})^2 \epsilon(a_{1}, k, p_{1},
          k_{1}) +
        4 (a_{2}.k_{1})\\&\times &\nonumber (k.p_{1})^2 (k.p_{2}) \epsilon(a_{1}, k, p_{1},
          k_{2}) -
        4 (a_{2}.k_{2}) (k.p_{1})^2 (k.p_{2})  \epsilon(a_{1}, k, p_{1},
          k_{2}) +
        4 (a_{2}.k_{1}) (k.p_{1}) (k.p_{2})^2 \epsilon(a_{1}, k, p_{1},
          k_{2}) -
        4 (a_{2}.k_{2}) (k.p_{1}) (k.p_{2})^2 \epsilon(a_{1}, k, p_{1},
          k_{2}) \\&+ &\nonumber
        4 (a_{2}.k_{1}) (k.p_{1})^2 (k.p_{2}) \epsilon(a_{1}, k, p_{2},
          k_{1}) -
        4 (a_{2}.k_{2}) (k.p_{1})^2 (k.p_{2}) \epsilon(a_{1}, k, p_{2},
          k_{1}) +
        4 (a_{2}.k_{1}) (k.p_{1}) (k.p_{2})^2  \epsilon(a_{1}, k, p_{2},
          k_{1}) -
        4 (a_{2}.k_{2}) (k.p_{1}) (k.p_{2})^2\\&\times &\nonumber \epsilon(a_{1}, k, p_{2},
          k_{1}) -
        4 (a_{2}.k_{1}) (k.p_{1})^2 (k.p_{2}) \epsilon(a_{1}, k, p_{2},
          k_{2}) +
        4 (a_{2}.k_{2}) (k.p_{1})^2 (k.p_{2}) \epsilon(a_{1}, k, p_{2},
          k_{2}) -
        4 (a_{2}.k_{1}) (k.p_{1}) (k.p_{2})^2 \epsilon(a_{1}, k, p_{2},
          k_{2}) +
        4 (a_{2}.k_{2})\\&\times &\nonumber (k.p_{1}) (k.p_{2})^2 \epsilon(a_{1}, k, p_{2},
          k_{2}) -
        4 (a_{1}.k_{1}) (k.p_{1}) (k.p_{2}) (k.k_{1}) \epsilon(a_{2}, k,
          p_{1}, p_{2}) +
        4 (a_{1}.k_{2}) (k.p_{1}) (k.p_{2}) (k.k_{1}) \epsilon(a_{2}, k,
          p_{1}, p_{2}) +
        4 (a_{1}.k_{1}) (k.p_{1}) (k.p_{2})\\&\times &\nonumber  (k.k_{2}) \epsilon(a_{2}, k,
          p_{1}, p_{2}) -
        4 (a_{1}.k_{2}) (k.p_{1}) (k.p_{2}) (k.k_{2}) \epsilon(a_{2}, k,
          p_{1}, p_{2}) +
        4 (a_{1}.k_{1}) (k.p_{1})^2 (k.p_{2}) \epsilon(a_{2}, k, p_{1},
          k_{1}) -
        4 (a_{1}.k_{2}) (k.p_{1})^2 (k.p_{2}) \epsilon(a_{2}, k, p_{1},
          k_{1}) \\&+ &\nonumber
        4 (a_{1}.k_{1}) (k.p_{1}) (k.p_{2})^2 \epsilon(a_{2}, k, p_{1},
          k_{1}) -
        4 (a_{1}.k_{2}) (k.p_{1}) (k.p_{2})^2  \epsilon(a_{2}, k, p_{1},
          k_{1}) -
        (a_{1}.p_{2}) (k.p_{1}) (k.p_{2}) (k.k_{1}) \epsilon(a_{2}, k, p_{1},
          k_{1}) +
        (a_{1}.p_{2}) (k.p_{2})^2 (k.k_{1})\\&\times &\nonumber \epsilon(a_{2}, k, p_{1},
          k_{1}) +
        (a_{1}.p_{2})  (k.p_{1}) (k.p_{2}) (k.k_{2}) \epsilon(a_{2}, k, p_{1},
          k_{1}) -
        (a_{1}.p_{2}) (k.p_{2})^2 (k.k_{2}) \epsilon(a_{2}, k, p_{1},
          k_{1}) -
        4 (a_{1}.k_{1}) (k.p_{1})^2 (k.p_{2}) \epsilon(a_{2}, k, p_{1},
          k_{2}) +
        4 (a_{1}.k_{2})\\&\times &\nonumber (k.p_{1})^2 (k.p_{2}) \epsilon(a_{2}, k, p_{1},
          k_{2}) -
        4 (a_{1}.k_{1}) (k.p_{1}) (k.p_{2})^2 \epsilon(a_{2}, k, p_{1},
          k_{2}) +
        4 (a_{1}.k_{2}) (k.p_{1}) (k.p_{2})^2 \epsilon(a_{2}, k, p_{1},
          k_{2}) +
        (a_{1}.p_{2}) (k.p_{1}) (k.p_{2}) (k.k_{1})\\&\times &\nonumber \epsilon(a_{2}, k, p_{1},
          k_{2}) -
        (a_{1}.p_{2}) (k.p_{2})^2 (k.k_{1}) \epsilon(a_{2}, k, p_{1},
          k_{2}) -
        (a_{1}.p_{2}) (k.p_{1}) (k.p_{2}) (k.k_{2}) \epsilon(a_{2}, k, p_{1},
          k_{2}) +
        (a_{1}.p_{2}) (k.p_{2})^2 (k.k_{2}) \epsilon(a_{2}, k, p_{1},
          k_{2}) -
        4 (a_{1}.k_{1})\\&\times &\nonumber (k.p_{1})^2 (k.p_{2}) \epsilon(a_{2}, k, p_{2},
          k_{1}) +
        4 (a_{1}.k_{2}) (k.p_{1})^2 (k.p_{2}) \epsilon(a_{2}, k, p_{2},
          k_{1}) -
        4 (a_{1}.k_{1}) (k.p_{1}) (k.p_{2})^2 \epsilon(a_{2}, k, p_{2},
          k_{1}) +
        4 (a_{1}.k_{2}) (k.p_{1}) (k.p_{2})^2 \epsilon(a_{2}, k, p_{2},
          k_{1}) \\&- &\nonumber
        (a_{1}.p_{1}) (k.p_{1})^2 (k.k_{1}) \epsilon(a_{2}, k, p_{2},
          k_{1}) +
        (a_{1}.p_{1}) (k.p_{1}) (k.p_{2}) (k.k_{1}) \epsilon(a_{2}, k, p_{2},
          k_{1}) +
        (a_{1}.p_{1}) (k.p_{1})^2 (k.k_{2}) \epsilon(a_{2}, k, p_{2},
          k_{1}) -
        (a_{1}.p_{1}) (k.p_{1}) (k.p_{2}) (k.k_{2})\\&\times &\nonumber \epsilon(a_{2}, k, p_{2},
          k_{1}) +
        (k.p_{1})  (4 (a_{1}.k_{1}) (k.p_{2}) ((k.p_{1}) + (k.p_{2})) - 4 (a_{1}.k_{2}) (k.p_{2}) ((k.p_{1}) + (k.p_{2})) +
           (a_{1}.p_{1}) ((k.p_{1}) - (k.p_{2})) ((k.k_{1}) - (k.k_{2}))) \epsilon(a_{2},
          k, p_{2}, k_{2})))\Big]
\end{eqnarray}
\normalsize

\tiny
\begin{eqnarray}
\Delta_{3}&=&\nonumber\dfrac{-e^2}{(2 (k.p_{1})^2 (k.p_{2})^2)}
    \Big[(-4 (k.p_{1}) (k.p_{2}) (-(a_{2}.k_{1})^2 g_{a}^{e^{2}} (k.p_{1}) (k.p_{2}) +
        2 (a_{2}.k_{1}) (a_{2}.k_{2}) g_{a}^{e^{2}} (k.p_{1}) (k.p_{2}) - (a_{2}.k_{2})^2 g_{a}^{e^{2}} (k.p_{1}) (k.p_{2}) -
        (a_{2}.k_{1})^2 g_{v}^{e^{2}} (k.p_{1})\\&\times &\nonumber (k.p_{2}) + 2 (a_{2}.k_{1}) (a_{2}.k_{2}) g_{v}^{e^{2}} (k.p_{1}) (k.p_{2}) -
        (a_{2}.k_{2})^2 g_{v}^{e^{2}} (k.p_{1}) (k.p_{2}) - (a_{1}.k_{1})^2 (g_{a}^{e^{2}} + g_{v}^{e^{2}}) (k.p_{1}) (k.p_{2}) -
        (a_{1}.k_{2})^2 (g_{a}^{e^{2}} + g_{v}^{e^{2}}) (k.p_{1}) (k.p_{2}) \\&- &\nonumber (a_{1}.p_{1}) (a_{1}.p_{2}) g_{a}^{e^{2}} (k.k_{1})^2 -
        (a_{1}.p_{1}) (a_{1}.p_{2}) g_{v}^{e^{2}} (k.k_{1})^2 +
        (a_{1}.k_{1}) (g_{a}^{e^{2}} +
           g_{v}^{e^{2}}) (2 (a_{1}.k_{2}) (k.p_{1}) (k.p_{2}) + ((a_{1}.p_{2}) (k.p_{1}) + (a_{1}.p_{1}) (k.p_{2})) ((k.k_{1}) \\&- &\nonumber
              (k.k_{2}))) -
        (a_{1}.k_{2}) (g_{a}^{e^{2}} + g_{v}^{e^{2}}) ((a_{1}.p_{2}) (k.p_{1}) + (a_{1}.p_{1}) (k.p_{2})) ((k.k_{1}) - (k.k_{2})) +
        2 (a_{1}.p_{1}) (a_{1}.p_{2}) g_{a}^{e^{2}} (k.k_{1}) (k.k_{2}) + 2 (a_{1}.p_{1}) (a_{1}.p_{2}) g_{v}^{e^{2}} (k.k_{1})\\&\times &\nonumber (k.k_{2}) -
        (a_{1}.p_{1}) (a_{1}.p_{2}) g_{a}^{e^{2}} (k.k_{2})^2 - (a_{1}.p_{1}) (a_{1}.p_{2}) g_{v}^{e^{2}} (k.k_{2})^2 +
        a^2 g_{a}^{e^{2}} (k.p_{1})^2 M_{A^{0}}^{2} + a^2 g_{v}^{e^{2}} (k.p_{1})^2 M_{A^{0}}^{2} +
        2 a^2 g_{a}^{e^{2}} (k.p_{1}) (k.p_{2}) M_{A^{0}}^{2} + 2 a^2 g_{v}^{e^{2}} \\&\times &\nonumber(k.p_{1}) (k.p_{2}) M_{A^{0}}^{2} +
        a^2 g_{a}^{e^{2}} (k.p_{2})^2 M_{A^{0}}^{2} + a^2 g_{v}^{e^{2}} (k.p_{2})^2 M_{A^{0}}^{2} -
        a^2 g_{a}^{e^{2}} (k.k_{1})^2 me^2 + a^2 g_{v}^{e^{2}} (k.k_{1})^2 me^2 +
        2 a^2 g_{a}^{e^{2}} (k.k_{1}) (k.k_{2}) me^2 - 2 a^2 g_{v}^{e^{2}} (k.k_{1})\\&\times &\nonumber (k.k_{2}) me^2 -
        a^2 g_{a}^{e^{2}} (k.k_{2})^2 me^2 + a^2 g_{v}^{e^{2}} (k.k_{2})^2 me^2 +
        a^2 g_{a}^{e^{2}} (k.k_{1})^2 (p_{1}.p_{2}) + a^2 g_{v}^{e^{2}} (k.k_{1})^2 (p_{1}.p_{2}) -
        2 a^2 g_{a}^{e^{2}} (k.k_{1}) (k.k_{2}) (p_{1}.p_{2}) - 2 a^2 g_{v}^{e^{2}} (k.k_{1})\\&\times &\nonumber (k.k_{2}) (p_{1}.p_{2}) +
        a^2 g_{a}^{e^{2}} (k.k_{2})^2 (p_{1}.p_{2}) + a^2 g_{v}^{e^{2}} (k.k_{2})^2 (p_{1}.p_{2}) -
        a^2 g_{a}^{e^{2}} (k.p_{1}) (k.k_{1}) (p_{1}.k_{1}) - a^2 g_{v}^{e^{2}} (k.p_{1}) (k.k_{1}) (p_{1}.k_{1}) -
        a^2 g_{a}^{e^{2}} (k.p_{2}) (k.k_{1}) (p_{1}.k_{1})\\&-&\nonumber a^2 g_{v}^{e^{2}} (k.p_{2}) (k.k_{1}) (p_{1}.k_{1}) +
        a^2 g_{a}^{e^{2}} (k.p_{1}) (k.k_{2}) (p_{1}.k_{1}) + a^2 g_{v}^{e^{2}} (k.p_{1}) (k.k_{2}) (p_{1}.k_{1}) +
        a^2 g_{a}^{e^{2}} (k.p_{2}) (k.k_{2}) (p_{1}.k_{1}) + a^2 g_{v}^{e^{2}} (k.p_{2}) (k.k_{2}) (p_{1}.k_{1}) +
        a^2 g_{a}^{e^{2}}\\&\times &\nonumber (k.p_{1}) (k.k_{1}) (p_{1}.k_{2}) + a^2 g_{v}^{e^{2}} (k.p_{1}) (k.k_{1}) (p_{1}.k_{2}) +
        a^2 g_{a}^{e^{2}} (k.p_{2}) (k.k_{1}) (p_{1}.k_{2}) + a^2 g_{v}^{e^{2}} (k.p_{2}) (k.k_{1}) (p_{1}.k_{2}) -
        a^2 g_{a}^{e^{2}} (k.p_{1}) (k.k_{2}) (p_{1}.k_{2}) - a^2 g_{v}^{e^{2}} \\&\times &\nonumber(k.p_{1}) (k.k_{2}) (p_{1}.k_{2}) -
        a^2 g_{a}^{e^{2}} (k.p_{2}) (k.k_{2}) (p_{1}.k_{2}) - a^2 g_{v}^{e^{2}} (k.p_{2}) (k.k_{2}) (p_{1}.k_{2}) -
        a^2 g_{a}^{e^{2}} (k.p_{1}) (k.k_{1}) (p_{2}.k_{1}) - a^2 g_{v}^{e^{2}} (k.p_{1}) (k.k_{1}) (p_{2}.k_{1}) -
        a^2 g_{a}^{e^{2}} \\&\times &\nonumber(k.p_{2}) (k.k_{1}) (p_{2}.k_{1}) - a^2 g_{v}^{e^{2}} (k.p_{2}) (k.k_{1}) (p_{2}.k_{1}) +
        a^2 g_{a}^{e^{2}} (k.p_{1}) (k.k_{2}) (p_{2}.k_{1}) + a^2 g_{v}^{e^{2}} (k.p_{1}) (k.k_{2}) (p_{2}.k_{1}) +
        a^2 g_{a}^{e^{2}} (k.p_{2}) (k.k_{2}) (p_{2}.k_{1}) + a^2 g_{v}^{e^{2}} \\&\times &\nonumber(k.p_{2}) (k.k_{2}) (p_{2}.k_{1}) +
        a^2 g_{a}^{e^{2}} (k.p_{1}) (k.k_{1}) (p_{2}.k_{2}) + a^2 g_{v}^{e^{2}} (k.p_{1}) (k.k_{1}) (p_{2}.k_{2}) +
        a^2 g_{a}^{e^{2}} (k.p_{2}) (k.k_{1}) (p_{2}.k_{2}) + a^2 g_{v}^{e^{2}} (k.p_{2}) (k.k_{1}) (p_{2}.k_{2}) -
        a^2 g_{a}^{e^{2}} \\&\times &\nonumber(k.p_{1}) (k.k_{2}) (p_{2}.k_{2}) - a^2 g_{v}^{e^{2}} (k.p_{1}) (k.k_{2}) (p_{2}.k_{2}) -
        a^2 g_{a}^{e^{2}} (k.p_{2}) (k.k_{2}) (p_{2}.k_{2}) - a^2 g_{v}^{e^{2}} (k.p_{2}) (k.k_{2}) (p_{2}.k_{2}) -
        a^2 (g_{a}^{e^{2}} + g_{v}^{e^{2}}) ((k.p_{1}) + (k.p_{2}))^2 (k_{1}.k_{2})) \\&+ &\nonumber
     g_{a}^{e} g_{v}^{e} (k.p_{2}) (4 (k.p_{1})^2 (-M_{A^{0}}^{2} + (k_{1}.k_{2})) +
        (k.p_{2}) (3 ((k.k_{1}) - (k.k_{2})) ((p_{2}.k_{1}) - (p_{2}.k_{2})) + 4 (k.p_{2}) (-M_{A^{0}}^{2} + (k_{1}.k_{2}))) +
        (k.p_{1}) (((k.k_{1}) - (k.k_{2}))\\&\times &\nonumber ((p_{2}.k_{1}) - (p_{2}.k_{2})) + 8 (k.p_{2}) (-M_{A^{0}}^{2} + (k_{1}.k_{2})))) \epsilon(
       a_{1}, a_{2}, k, p_{1}) +
     g_{a}^{e} g_{v}^{e} (k.p_{1}) ((k.p_{2}) (-((k.k_{1}) - (k.k_{2})) ((p_{1}.k_{1}) - (p_{1}.k_{2})) +
           4 (k.p_{2}) (M_{A^{0}}^{2} \\&- &\nonumber (k_{1}.k_{2}))) +
        (k.p_{1}) (-3 ((k.k_{1}) - (k.k_{2})) ((p_{1}.k_{1}) - (p_{1}.k_{2})) + 8 (k.p_{2}) (M_{A^{0}}^{2} - (k_{1}.k_{2}))) +
        4 (k.p_{1})^2 (M_{A^{0}}^{2} - (k_{1}.k_{2}))) \epsilon(a_{1}, a_{2},
       k, p_{2}) +
     g_{a}^{e} g_{v}^{e}\\&\times &\nonumber (-((k.p_{1}) - (k.p_{2})) ((k.p_{1}) + (k.p_{2})) ((k.k_{1}) - (k.k_{2})) (p_{1}.p_{2}) \epsilon(
          a_{1}, a_{2}, k,
          k_{1}) + ((k.p_{1}) - (k.p_{2})) ((k.p_{1}) + (k.p_{2})) ((k.k_{1}) -
           (k.k_{2})) (p_{1}.p_{2}) \\&\times &\nonumber\epsilon(a_{1}, a_{2}, k,
          k_{2}) -
        3 (k.p_{1})^2 (k.k_{1})^2 \epsilon(a_{1}, a_{2}, p_{1},
          p_{2}) +
        6 (k.p_{1}) (k.p_{2}) (k.k_{1})^2 \epsilon(a_{1}, a_{2}, p_{1},
          p_{2}) -
        3 (k.p_{2})^2 (k.k_{1})^2 \epsilon(a_{1}, a_{2}, p_{1},
          p_{2}) +
        6 (k.p_{1})^2 (k.k_{1}) (k.k_{2})\\&\times &\nonumber \epsilon(a_{1}, a_{2}, p_{1},
          p_{2}) -
        12 (k.p_{1}) (k.p_{2}) (k.k_{1}) (k.k_{2}) \epsilon(a_{1}, a_{2},
          p_{1}, p_{2}) +
        6 (k.p_{2})^2 (k.k_{1}) (k.k_{2}) \epsilon(a_{1}, a_{2}, p_{1},
          p_{2}) -
        3 (k.p_{1})^2 (k.k_{2})^2 \epsilon(a_{1}, a_{2}, p_{1},
          p_{2}) +
        6 (k.p_{1}) \\&\times &\nonumber(k.p_{2}) (k.k_{2})^2 \epsilon(a_{1}, a_{2}, p_{1},
          p_{2}) -
        3 (k.p_{2})^2 (k.k_{2})^2 \epsilon(a_{1}, a_{2}, p_{1},
          p_{2}) -
        (k.p_{1})^2 (k.p_{2}) (k.k_{1}) \epsilon(a_{1}, a_{2}, p_{1},
          k_{1}) -
        5 (k.p_{1}) (k.p_{2})^2 (k.k_{1}) \epsilon(a_{1}, a_{2}, p_{1},
          k_{1}) \\&+ &\nonumber
        2 (k.p_{2})^3 (k.k_{1}) \epsilon(a_{1}, a_{2}, p_{1},
          k_{1}) +
        (k.p_{1})^2 (k.p_{2}) (k.k_{2}) \epsilon(a_{1}, a_{2}, p_{1},
          k_{1}) +
        5 (k.p_{1}) (k.p_{2})^2 (k.k_{2}) \epsilon(a_{1}, a_{2}, p_{1},
          k_{1}) -
        2 (k.p_{2})^3 (k.k_{2}) \epsilon(a_{1}, a_{2}, p_{1},
          k_{1}) \\&+ &\nonumber
        (k.p_{1})^2 (k.p_{2}) (k.k_{1}) \epsilon(a_{1}, a_{2}, p_{1},
          k_{2}) +
        5 (k.p_{1}) (k.p_{2})^2 (k.k_{1}) \epsilon(a_{1}, a_{2}, p_{1},
          k_{2}) -
        2 (k.p_{2})^3 (k.k_{1}) \epsilon(a_{1}, a_{2}, p_{1},
          k_{2}) -
        (k.p_{1})^2 (k.p_{2}) (k.k_{2}) \epsilon(a_{1}, a_{2}, p_{1},
          k_{2}) \\&- &\nonumber
        5 (k.p_{1}) (k.p_{2})^2 (k.k_{2}) \epsilon(a_{1}, a_{2}, p_{1},
          k_{2}) +
        2 (k.p_{2})^3 (k.k_{2}) \epsilon(a_{1}, a_{2}, p_{1},
          k_{2}) -
        2 (k.p_{1})^3 (k.k_{1}) \epsilon(a_{1}, a_{2}, p_{2},
          k_{1}) +
        5 (k.p_{1})^2 (k.p_{2}) (k.k_{1}) \epsilon(a_{1}, a_{2}, p_{2},
          k_{1}) \\&+ &\nonumber
        (k.p_{1}) (k.p_{2})^2 (k.k_{1}) \epsilon(a_{1}, a_{2}, p_{2},
          k_{1}) +
        2 (k.p_{1})^3 (k.k_{2}) \epsilon(a_{1}, a_{2}, p_{2},
          k_{1}) -
        5 (k.p_{1})^2 (k.p_{2}) (k.k_{2}) \epsilon(a_{1}, a_{2}, p_{2},
          k_{1}) -
        (k.p_{1}) (k.p_{2})^2 (k.k_{2}) \epsilon(a_{1}, a_{2}, p_{2},
          k_{1}) \\&+ &\nonumber
        2 (k.p_{1})^3 (k.k_{1}) \epsilon(a_{1}, a_{2}, p_{2},
          k_{2}) -
        5 (k.p_{1})^2 (k.p_{2}) (k.k_{1}) \epsilon(a_{1}, a_{2}, p_{2},
          k_{2}) -
        (k.p_{1}) (k.p_{2})^2 (k.k_{1}) \epsilon(a_{1}, a_{2}, p_{2},
          k_{2}) -
        2 (k.p_{1})^3 (k.k_{2}) \epsilon(a_{1}, a_{2}, p_{2},
          k_{2})\\&+ &\nonumber
        5 (k.p_{1})^2 (k.p_{2}) (k.k_{2}) \epsilon(a_{1}, a_{2}, p_{2},
          k_{2}) +
        (k.p_{1}) (k.p_{2})^2 (k.k_{2}) \epsilon(a_{1}, a_{2}, p_{2},
          k_{2}) -
        4 (a_{2}.k_{1}) (k.p_{1}) (k.p_{2}) (k.k_{1}) \epsilon(a_{1}, k,
          p_{1}, p_{2}) +
        4 (a_{2}.k_{2}) (k.p_{1}) (k.p_{2}) (k.k_{1})\\&\times &\nonumber \epsilon(a_{1}, k,
          p_{1}, p_{2}) +
        4 (a_{2}.k_{1}) (k.p_{1}) (k.p_{2}) (k.k_{2}) \epsilon(a_{1}, k,
          p_{1}, p_{2}) -
        4 (a_{2}.k_{2}) (k.p_{1}) (k.p_{2}) (k.k_{2}) \epsilon(a_{1}, k,
          p_{1}, p_{2}) +
        4 (a_{2}.k_{1}) (k.p_{1})^2 (k.p_{2}) \epsilon(a_{1}, k, p_{1},
          k_{1}) \\&- &\nonumber
        4 (a_{2}.k_{2}) (k.p_{1})^2 (k.p_{2}) \epsilon(a_{1}, k, p_{1},
          k_{1}) +
        4 (a_{2}.k_{1}) (k.p_{1}) (k.p_{2})^2 \epsilon(a_{1}, k, p_{1},
          k_{1}) -
        4 (a_{2}.k_{2}) (k.p_{1}) (k.p_{2})^2 \epsilon(a_{1}, k, p_{1},
          k_{1}) -
        4 (a_{2}.k_{1}) (k.p_{1})^2 (k.p_{2})\\&\times &\nonumber \epsilon(a_{1}, k, p_{1},
          k_{2}) +
        4 (a_{2}.k_{2}) (k.p_{1})^2 (k.p_{2}) \epsilon(a_{1}, k, p_{1},
          k_{2}) -
        4 (a_{2}.k_{1}) (k.p_{1}) (k.p_{2})^2 \epsilon(a_{1}, k, p_{1},
          k_{2}) +
        4 (a_{2}.k_{2}) (k.p_{1}) (k.p_{2})^2 \epsilon(a_{1}, k, p_{1},
          k_{2}) -
        4 (a_{2}.k_{1})\\&\times &\nonumber (k.p_{1})^2 (k.p_{2}) \epsilon(a_{1}, k, p_{2},
          k_{1}) +
        4 (a_{2}.k_{2}) (k.p_{1})^2 (k.p_{2}) \epsilon(a_{1}, k, p_{2},
          k_{1}) -
        4 (a_{2}.k_{1}) (k.p_{1}) (k.p_{2})^2 \epsilon(a_{1}, k, p_{2},
          k_{1}) +
        4 (a_{2}.k_{2}) (k.p_{1}) (k.p_{2})^2 \epsilon(a_{1}, k, p_{2},
          k_{1}) \\&+ &\nonumber
        4 (a_{2}.k_{1}) (k.p_{1})^2 (k.p_{2}) \epsilon(a_{1}, k, p_{2},
          k_{2}) -
        4 (a_{2}.k_{2}) (k.p_{1})^2 (k.p_{2}) \epsilon(a_{1}, k, p_{2},
          k_{2}) +
        4 (a_{2}.k_{1}) (k.p_{1}) (k.p_{2})^2 \epsilon(a_{1}, k, p_{2},
          k_{2}) -
        4 (a_{2}.k_{2}) (k.p_{1}) (k.p_{2})^2\\&\times &\nonumber \epsilon(a_{1}, k, p_{2},
          k_{2}) +
        4 (a_{1}.k_{1}) (k.p_{1}) (k.p_{2}) (k.k_{1}) \epsilon(a_{2}, k,
          p_{1}, p_{2}) -
        4 (a_{1}.k_{2}) (k.p_{1}) (k.p_{2}) (k.k_{1}) \epsilon(a_{2}, k,
          p_{1}, p_{2}) -
        4 (a_{1}.k_{1}) (k.p_{1}) (k.p_{2}) (k.k_{2}) \epsilon(a_{2}, k,
          p_{1}, p_{2}) \\&+ &\nonumber
        4 (a_{1}.k_{2}) (k.p_{1}) (k.p_{2}) (k.k_{2}) \epsilon(a_{2}, k,
          p_{1}, p_{2}) -
        4 (a_{1}.k_{1}) (k.p_{1})^2 (k.p_{2}) \epsilon(a_{2}, k, p_{1},
          k_{1}) +
        4 (a_{1}.k_{2}) (k.p_{1})^2 (k.p_{2}) \epsilon(a_{2}, k, p_{1},
          k_{1}) -
        4 (a_{1}.k_{1}) (k.p_{1}) (k.p_{2})^2\\&\times &\nonumber \epsilon(a_{2}, k, p_{1},
          k_{1}) +
        4 (a_{1}.k_{2}) (k.p_{1}) (k.p_{2})^2 \epsilon(a_{2}, k, p_{1},
          k_{1}) +
        (a_{1}.p_{2}) (k.p_{1}) (k.p_{2}) (k.k_{1}) \epsilon(a_{2}, k, p_{1},
          k_{1}) -
        (a_{1}.p_{2}) (k.p_{2})^2 (k.k_{1}) \epsilon(a_{2}, k, p_{1},
          k_{1}) -
        (a_{1}.p_{2})\\&\times &\nonumber (k.p_{1}) (k.p_{2}) (k.k_{2}) \epsilon(a_{2}, k, p_{1},
          k_{1}) +
        (a_{1}.p_{2}) (k.p_{2})^2 (k.k_{2}) \epsilon(a_{2}, k, p_{1},
          k_{1}) +
        4 (a_{1}.k_{1}) (k.p_{1})^2 (k.p_{2}) \epsilon(a_{2}, k, p_{1},
          k_{2}) -
        4 (a_{1}.k_{2}) (k.p_{1})^2 (k.p_{2})\\&\times &\nonumber \epsilon(a_{2}, k, p_{1},
          k_{2}) +
        4 (a_{1}.k_{1}) (k.p_{1}) (k.p_{2})^2 \epsilon(a_{2}, k, p_{1},
          k_{2}) -
        4 (a_{1}.k_{2}) (k.p_{1}) (k.p_{2})^2 \epsilon(a_{2}, k, p_{1},
          k_{2}) -
        (a_{1}.p_{2}) (k.p_{1}) (k.p_{2}) (k.k_{1}) \epsilon(a_{2}, k, p_{1},
          k_{2}) +
        (a_{1}.p_{2})\\&\times &\nonumber (k.p_{2})^2 (k.k_{1}) \epsilon(a_{2}, k, p_{1},
          k_{2}) +
        (a_{1}.p_{2}) (k.p_{1}) (k.p_{2}) (k.k_{2}) \epsilon(a_{2}, k, p_{1},
          k_{2}) -
        (a_{1}.p_{2}) (k.p_{2})^2 (k.k_{2}) \epsilon(a_{2}, k, p_{1},
          k_{2}) +
        4 (a_{1}.k_{1}) (k.p_{1})^2 (k.p_{2}) \epsilon(a_{2}, k, p_{2},
          k_{1}) \\&- &\nonumber
        4 (a_{1}.k_{2}) (k.p_{1})^2 (k.p_{2}) \epsilon(a_{2}, k, p_{2},
          k_{1}) +
        4 (a_{1}.k_{1}) (k.p_{1}) (k.p_{2})^2 \epsilon(a_{2}, k, p_{2},
          k_{1}) -
        4 (a_{1}.k_{2}) (k.p_{1}) (k.p_{2})^2 \epsilon(a_{2}, k, p_{2},
          k_{1}) +
        (a_{1}.p_{1}) (k.p_{1})^2 (k.k_{1})\\&\times &\nonumber \epsilon(a_{2}, k, p_{2},
          k_{1}) -
        (a_{1}.p_{1}) (k.p_{1}) (k.p_{2}) (k.k_{1}) \epsilon(a_{2}, k, p_{2},
          k_{1}) -
        (a_{1}.p_{1}) (k.p_{1})^2 (k.k_{2}) \epsilon(a_{2}, k, p_{2},
          k_{1}) +
        (a_{1}.p_{1}) (k.p_{1}) (k.p_{2}) (k.k_{2}) \epsilon(a_{2}, k, p_{2},
          k_{1}) +
        (k.p_{1})\\&\times &\nonumber (-4 (a_{1}.k_{1}) (k.p_{2}) ((k.p_{1}) + (k.p_{2})) + 4 (a_{1}.k_{2}) (k.p_{2}) ((k.p_{1}) + (k.p_{2})) -
           (a_{1}.p_{1}) ((k.p_{1}) - (k.p_{2})) ((k.k_{1}) - (k.k_{2}))) \epsilon(a_{2},
          k, p_{2}, k_{2})))\Big]
\end{eqnarray}
\normalsize
\tiny
\begin{eqnarray}
\Delta_{4}&=&\nonumber \dfrac{2\,e}{((k.p_{1})^2 (k.p_{2})^2)}\Big[
   ((g_{a}^{e^{2}} +
       g_{v}^{e^{2}}) (k.p_{1}) (k.p_{2}) (a^2 e^2 (((a_{1}.k_{1}) - (a_{1}.k_{2})) ((k.p_{1}) - (k.p_{2})) - ((a_{1}.p_{1}) -
             (a_{1}.p_{2})) ((k.k_{1}) - (k.k_{2}))) ((k.k_{1}) - (k.k_{2})) \\&+ &\nonumber
       2 (((a_{1}.k_{1}) - (a_{1}.k_{2})) (k.p_{1}) (k.p_{2}) ((p_{1}.k_{1}) - (p_{1}.k_{2}) - (p_{2}.k_{1}) + (p_{2}.k_{2})) +
          (a_{1}.p_{1}) (k.p_{2}) (((k.k_{1}) - (k.k_{2})) ((p_{2}.k_{1}) - (p_{2}.k_{2})) - ((k.p_{1}) + (k.p_{2}))\\&\times &\nonumber (M_{A^{0}}^{2} -
                (k_{1}.k_{2}))) +
          (a_{1}.p_{2}) (k.p_{1}) (-((k.k_{1}) - (k.k_{2})) ((p_{1}.k_{1}) - (p_{1}.k_{2})) + ((k.p_{1}) + (k.p_{2})) (M_{A^{0}}^{2} -
                (k_{1}.k_{2}))))) -
    a^2 e^2 g_{a}^{e} g_{v}^{e} ((k.p_{1}) - (k.p_{2}))\\&\times &\nonumber ((k.k_{1}) - (k.k_{2}))^2 \epsilon(a_{2},
      k, p_{1}, p_{2}) -
    g_{a}^{e} g_{v}^{e} (k.p_{2}) ((k.p_{1}) + (k.p_{2})) (a^2 e^2 ((k.k_{1}) - (k.k_{2})) +
       2 (k.p_{1}) ((p_{2}.k_{1}) - (p_{2}.k_{2}))) \epsilon(a_{2}, k,
      p_{1}, k_{1}) +
    g_{a}^{e} g_{v}^{e} ((k.p_{2})\\&\times &\nonumber ((k.p_{1}) + (k.p_{2})) (a^2 e^2 ((k.k_{1}) - (k.k_{2})) +
          2 (k.p_{1}) ((p_{2}.k_{1}) - (p_{2}.k_{2}))) \epsilon(a_{2}, k,
         p_{1}, k_{2}) +
       (k.p_{1}) (-((k.p_{1}) + (k.p_{2})) (a^2 e^2 ((k.k_{1}) - (k.k_{2})) +
             2 (k.p_{2}) \\&\times &\nonumber((p_{1}.k_{1}) - (p_{1}.k_{2}))) \epsilon(a_{2}, k,
            p_{2},
            k_{1}) + ((k.p_{1}) + (k.p_{2})) (a^2 e^2 ((k.k_{1}) - (k.k_{2})) +
             2 (k.p_{2}) ((p_{1}.k_{1}) - (p_{1}.k_{2}))) \epsilon(a_{2}, k,
            p_{2}, k_{2}) + 2 ((k.p_{1}) -
             (k.p_{2}))\\&\times &\nonumber (k.p_{2}) (-((k.k_{1}) - (k.k_{2})) (\epsilon(a_{2}, p_{1},
                 p_{2}, k_{1}) -
                \epsilon(a_{2}, p_{1}, p_{2},
                 k_{2})) + ((a_{2}.k_{1}) -
                (a_{2}.k_{2})) (\epsilon(k, p_{1}, p_{2},
                 k_{1}) -
                \epsilon(k, p_{1}, p_{2},
                 k_{2}))))))\Big]
\end{eqnarray}
\normalsize
\tiny
\begin{eqnarray}
\Delta_{5}&=&\nonumber\dfrac{2\,e}{((k.p_{1})^2 (k.p_{2})^2)}\Big[
   ((g_{a}^{e^{2}} +
       g_{v}^{e^{2}}) (k.p_{1}) (k.p_{2}) (a^2 e^2 (((a_{1}.k_{1}) - (a_{1}.k_{2})) ((k.p_{1}) - (k.p_{2})) - ((a_{1}.p_{1}) -
             (a_{1}.p_{2})) ((k.k_{1}) - (k.k_{2}))) ((k.k_{1}) - (k.k_{2})) \\&+ &\nonumber
       2 (((a_{1}.k_{1}) - (a_{1}.k_{2})) (k.p_{1}) (k.p_{2}) ((p_{1}.k_{1}) - (p_{1}.k_{2}) - (p_{2}.k_{1}) + (p_{2}.k_{2})) +
          (a_{1}.p_{1}) (k.p_{2}) (((k.k_{1}) - (k.k_{2})) ((p_{2}.k_{1}) - (p_{2}.k_{2})) - ((k.p_{1}) \\&+ &\nonumber(k.p_{2})) (M_{A^{0}}^{2} -
                (k_{1}.k_{2}))) +
          (a_{1}.p_{2}) (k.p_{1}) (-((k.k_{1}) - (k.k_{2})) ((p_{1}.k_{1}) - (p_{1}.k_{2})) + ((k.p_{1}) + (k.p_{2})) (M_{A^{0}}^{2} -
                (k_{1}.k_{2}))))) +
    a^2 e^2 g_{a}^{e} g_{v}^{e} ((k.p_{1}) \\&- &\nonumber (k.p_{2})) ((k.k_{1}) - (k.k_{2}))^2 \epsilon(a_{2},
      k, p_{1}, p_{2}) +
    g_{a}^{e} g_{v}^{e} (k.p_{2}) ((k.p_{1}) + (k.p_{2})) (a^2 e^2 ((k.k_{1}) - (k.k_{2})) +
       2 (k.p_{1}) ((p_{2}.k_{1}) - (p_{2}.k_{2}))) \epsilon(a_{2}, k,
      p_{1}, k_{1}) \\&+ &\nonumber
    g_{a}^{e} g_{v}^{e} (-(k.p_{2}) ((k.p_{1}) + (k.p_{2})) (a^2 e^2 ((k.k_{1}) - (k.k_{2})) +
          2 (k.p_{1}) ((p_{2}.k_{1}) - (p_{2}.k_{2}))) \epsilon(a_{2}, k,
         p_{1}, k_{2}) +
       (k.p_{1}) (((k.p_{1}) + (k.p_{2})) (a^2 e^2 ((k.k_{1}) \\&- &\nonumber(k.k_{2})) +
             2 (k.p_{2}) ((p_{1}.k_{1}) - (p_{1}.k_{2}))) \epsilon(a_{2}, k,
            p_{2},
            k_{1}) - ((k.p_{1}) + (k.p_{2})) (a^2 e^2 ((k.k_{1}) - (k.k_{2})) +
             2 (k.p_{2}) ((p_{1}.k_{1}) - (p_{1}.k_{2}))) \epsilon(a_{2}, k,
            p_{2}, k_{2}) \\&+ &\nonumber 2 ((k.p_{1}) -
             (k.p_{2})) (k.p_{2}) (((k.k_{1}) -
                (k.k_{2})) (\epsilon(a_{2}, p_{1}, p_{2},
                 k_{1}) -
                \epsilon(a_{2}, p_{1}, p_{2},
                 k_{2})) - ((a_{2}.k_{1}) -
                (a_{2}.k_{2})) (\epsilon(k, p_{1}, p_{2},
                 k_{1}) -
                \epsilon(k, p_{1}, p_{2},
                 k_{2}))))))\Big]
\end{eqnarray}
\normalsize
\tiny
\begin{eqnarray}
\Delta_{6}&=&\nonumber\dfrac{4 e^2}{((k.p_{1}) (k.p_{2}))}
    \Big[(g_{a}^{e^{2}} + g_{v}^{e^{2}}) ((a_{1}.k_{1})^2 (k.p_{1}) (k.p_{2}) +
     (a_{1}.k_{2})^2 (k.p_{1}) (k.p_{2}) - ((a_{2}.k_{1}) - (a_{2}.k_{2}))^2 (k.p_{1}) (k.p_{2}) +
     (a_{1}.k_{1}) (-2 (a_{1}.k_{2}) (k.p_{1})\\&\times &\nonumber (k.p_{2}) - ((a_{1}.p_{2}) (k.p_{1}) + (a_{1}.p_{1}) (k.p_{2})) ((k.k_{1}) - (k.k_{2}))) +
     (a_{1}.k_{2}) ((a_{1}.p_{2}) (k.p_{1}) + (a_{1}.p_{1}) (k.p_{2})) ((k.k_{1}) - (k.k_{2})) + (a_{1}.p_{1})\\&\times &\nonumber (a_{1}.p_{2}) ((k.k_{1}) - (k.k_{2}))^2)\Big]
\end{eqnarray}

\end{document}